%% file: threecirc5.tex
\documentclass[pre,twocolumn, showpacs, preprintnumbers, amsmath,
amssymb]{revtex4-1}
\usepackage[pdftex]{graphicx}
\usepackage{hyperref}
\usepackage{amssymb}
\usepackage{amsmath}

\usepackage{dcolumn}
\usepackage{bm}

\begin{document}
\title{A Free Energy Landscape for Cage Breaking of Three Hard Disks}
\author{Gary L.~Hunter}
\author{Eric R.~Weeks}
\address{Department of  Physics, Emory
University, Atlanta, GA 30322}
\begin{abstract}
We investigate cage breaking in dense hard disk systems using a model of three Brownian disks confined within a circular corral.  This system has a six-dimensional configuration space, but can be equivalently thought to explore a symmetric one-dimensional free energy landscape containing two energy minima separated by an energy barrier. The exact free energy landscape can be calculated as a function of system size.  Results of simulations show the average time between cage breaking events follows an Arrhenius scaling when the energy barrier is large.  We also discuss some of the consequences of using a one-dimensional representation to understand dynamics in a multi-dimensional space, such as diffusion acquiring spatial dependence and discontinuities in spatial derivatives of free energy.

\end{abstract}
\maketitle

\section{Introduction} 
\label{sec:intro}

The concept of a potential or free energy landscape for condensed matter systems is an appealing one~\cite{goldstein1969,stillinger1995,sciortino05}, and its use to  understand the glass transitions in various materials has received considerable attention~\cite{sastry98,dasgupta99,dasgupta00,schroder00, sastry01,grigera02,kim2003,vogel04, schweizerjcp,bonn07}.  Each configuration of a system can be assigned an energy based on how individual elements interact, and the set of all configurational energies then constitutes an energy surface or ``landscape''.  The temporal evolution of a system can be thought of as an exploration of various topographical features on the energy surface, with relaxation events understood as motions between adjacent energy minima  \cite{kauzmann48,stillinger1995,debenedetti2001nat,walesbook}.  As the glass transition is approached, the system finds itself in progressively deeper local minima on the landscape until thermal energy is unable to excite the system over a barrier and into a lower energy state.

A complete energy landscape accounts for all degrees of freedom within a system, each of which contributes a single dimension to the configuration space.  For many body systems then, the complete energy landscape can be extraordinarily complex.  A central problem in using the energy landscape approach is minimizing the number of ``reaction coordinates'' (likewise ``order parameters'') while still adequately describing the surface~\cite{frenkel92,frenkel96,frenkel97,heuer97,walesbook,krivov04,altis08,liwo09,bevan11}.  Studies often choose one or two coordinates of interest and examine how the system evolves on representative 1D or 2D landscapes, though in some cases such low dimensional projections may  introduce artifacts or occlude important features of the landscape~\cite{liwo09,altis08,krivov04}.

These complexities aside, the typical picture of an energy landscape is one of a surface containing hills and valleys.  However, for the case of a purely hard core potential, such as that of hard disks or hard spheres, this picture is incorrect.  In a hard disk or hard sphere system, the Helmholtz free energy is governed entirely by entropy, 
\begin{equation}\label{eqn:fts}
F = -TS,
\end{equation}
\noindent to within an additive constant  \cite{lowenhardspheres,donev07}.  As all allowable configurations have identically zero potential energy, they are all equiprobable.  Thus, in terms of the total configuration space, the energy landscape is completely flat. Upon projection into a lower dimensional space, however, hills and valleys related to entropic minima and maxima can arise~\cite{speedytwodiscs,dasgupta99,speedy1999discs,dasgupta00,schweizerjcp,schweizerjpcb}.

\begin{figure}\includegraphics[width=0.48\textwidth]{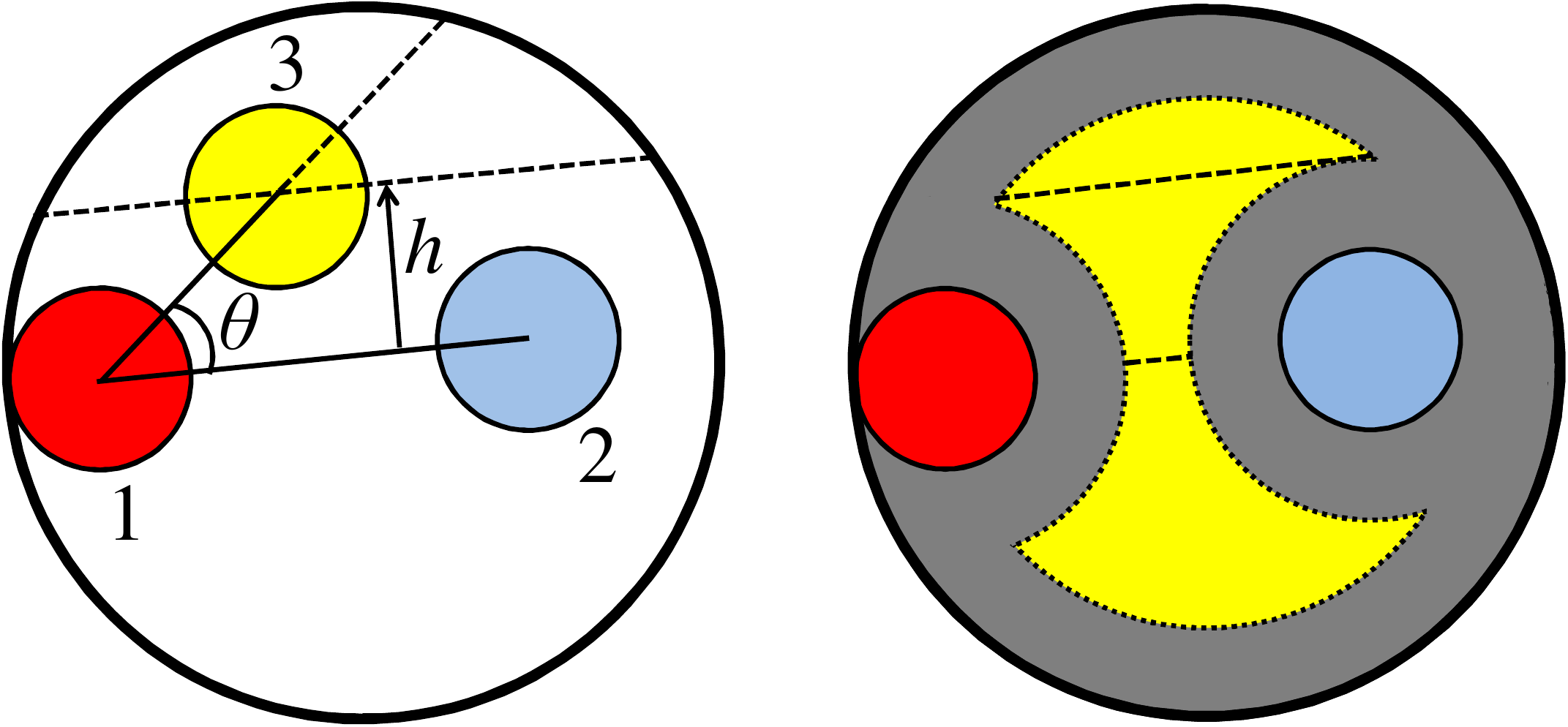}
 \caption{(Color online) (a). Three distinguishable hard disks are confined within a circular corral of fixed size. The variables $h$ and $\theta$ are macrostate variables, to be defined later.  Dashed lines are lines of constant $h$ and $\theta$.  (b). A 2D slice through the 6D configuration space with constant $(\vec{r}_1,\vec{r}_2)$.  The yellow (gray) region is accessible to the center of disk 3, whereas the dark gray region is energetically forbidden. Upper and lower dashed lines represent possible system macrostates at fixed $h$, where the of the line length determines the entropy of the state.} \label{fig:schematic}
\end{figure}

\section{Model System}\label{sec:model}

Consider a system of  $N = 3$ hard disks confined within a hard circular corral, as illustrated in Fig.~\ref{fig:schematic}(a).  The configuration space of this system is six-dimensional (6D) and is completely described by all allowable combinations of $(\vec{r}_1,\vec{r}_2,\vec{r}_3)$.  However, as stated previously, the energy landscape is flat because each configuration is as probable as any other.  Now, we consider a 2D slice though the landscape examining all configurations of $\vec{r}_3$ while $(\vec{r}_1,\vec{r}_2)$ are held at constant values, such as shown in Fig.~\ref{fig:schematic}(b). All allowed configurations are again equiprobable and are indicated by the light gray region (yellow online). The landscape in this scenario remains flat.  However, there are more configurations where disk 3 exists along the upper dashed line than on the line between disks 1 and 2. If we interpret these two lines as macrostates of the system, the entropy of the upper line is larger and therefore the free energy in Eqn.~(\ref{eqn:fts}) is lower.  By considering all macrostates, that is all lines parallel to those in Fig.~\ref{fig:schematic}(b), we obtain a 1D representative landscape with variations in free energy.

We can immediately relate these ideas to a dynamic version of the system in Fig.~\ref{fig:schematic}(a), where all disks wander around the corral by means of Brownian motion.  Here, we may ignore degrees of freedom associated with momentum and focus only on the six spatial components.  As the system explores 6D configuration space, disk 3 spends more time in configurations with high entropy on the 1D landscape, such as along the upper dashed line in Fig.~\ref{fig:schematic}(b).   To transition into the lower cusped region, it must pass through an entropic bottleneck.  Of course, this behavior is not unique to disk 3 but applies to all disks in our system.  Hence, the transition of any single disk corresponds to the entire system crossing a free energy barrier.  

Prior to a barrier crossing event, the motion of each disk is localized by the presence of other disks and the wall.  As with individual molecules in dense liquids or particles in a dense colloidal suspension, the disks can then be described as ``caged''~\cite{berne66, sjogren80,wahnstrom82, gotze1992rpp,doliwa1998prl,weeks02sub,weeks02,schweizerjcp,kvlee2004,sirono2011,wales2008,ohern2008}.  Deviations from strongly localized behavior, such as those during a crossing event, are considered ``cage breaking"~\cite{alder63, adam1965jcp,donati1998prl,marcus1999pre,doliwa2000pre, scheidler02,kvlee2004, sirono2011}.  One can consider both caging and cage breaking from the point of view of an energy landscape, where caging is the motion of the system around some local minimum and cage breaking is the relaxation over an energy barrier~\cite{schweizerjpcb,ediger2000arpc,richert02,sillescu1999,odagaki06,yoshidome07,yoshidome08}.

In the spirit of \cite{alder63,speedytwodiscs, speedy1999discs,heuer00,ohern2008}, we introduce the minimal model  system illustrated in Fig.~\ref{fig:schematic}(a) in order to explore the relationship between caging and the free energy landscape of a system with hard core potentials.  Here,  a cage breaking event can be simply described as one disk passing between the other two.  One can imagine for a very large corral, the motions of any disk would only rarely be influenced by the other disks or the boundary, and so dynamics would be similar to those in a dilute colloidal suspension.   However, this picture changes for smaller corral sizes, or similarly higher packing fractions, where interactions between the disks or with the wall are more frequent and dynamically restrictive. Though somewhat contrived, these purely geometric constraints are of similar character to those encountered by real particles in a densely packed system, such as colloidal supercooled liquids or glasses \cite{segrepusey97,pusey2008jpcm,cianci2006ssc,vanBlaaderen1995}.

\section{Simulation Details}
\label{sec:sim}

Three hard-disks of radius $r = d/2 = 1$ are confined to a hard-walled, circular corral of radius $R_C = 3 + \epsilon$.  Choosing $r = 1$, we note that all lengths are by definition in units of a particle radius.  The minimum system size that permits cage breaking occurs at $\epsilon = 0$, where the three particles exactly fit across the diameter of the corral.  During each simulational run, we fix the value of $\epsilon$ and allow particles to execute Brownian motion, described below.  Each run consists of $10^8$ Monte Carlo steps (mcs) for a particular value of $\epsilon$, during which each particle has the opportunity to make a displacement.  Values of $\epsilon$ are chosen within a range of $\epsilon \in [0.037,10.0]$.

To simulate Brownian dynamics, displacements for each particle at every time step in the $x-$ and $y-$directions are sampled from a Gaussian distribution with variance $\sigma^2 = 2D = 1 \times10^{-3}$. A displacement is accepted if it does not result in particle-particle or particle-wall overlap, otherwise the offending particle remains fixed for that time step.  This results in an RMS displacement of approximately $\approx 2\%$ of a particle diameter at each time step.  The value of $D$, and therefore the temperature, is constant across all simulations.  The order in which particles are sampled is randomized at each time step and satisfies detailed balance~\cite{frenkelbook}.   For the stated value of $D$, the fraction of accepted displacements range from $0.954$ at $\epsilon = 0.037$ to $0.997$ at $\epsilon = 10.0$.


\section{Dynamics in 1D}
\label{sec:dyn}

\begin{figure}\includegraphics[width=0.48\textwidth]{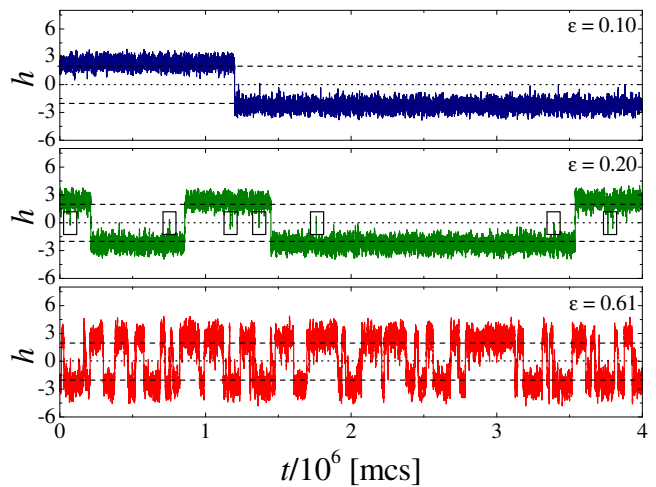}
 \caption{Trajectories through $h$-space for (top-bottom): $\epsilon = 0.10, \ 0.20, \ {\rm and}\ 0.61$.  Dotted lines pass through $h = 0$ and dashed lines pass through $h = \pm 2$.  The length of time between crossing events is clearly a sensitive function of $\epsilon$.  Small fluctuations about $h=0$ (such as those in the boxed regions for $\epsilon = 0.20$) are not considered as true cage breaking events.} \label{fig:traj}
\end{figure}
To study the dynamics of this system, we first simplify the 6D configuration space by projecting down to one dimension.  We define a macrostate variable $h$, shown in Fig.~\ref{fig:schematic}(a), as the distance of the center of disk 3 normal to a line segment drawn from the center of disk 1 to that of disk 2.  Put another way, if a line drawn from the center of disk 1 to that of disk 2 defines the positive $x$-axis of a coordinate system, $h$ is given by the $y$-coordinate of disk 3.  Therefore, $h$ can take positive or negative values, which from geometry are limited to the range $[-h_{\rm max},h_{\rm max}]$, where $h_{\rm max} = 2+\epsilon + \sqrt{3 + 4\epsilon + \epsilon^2}$. This definition allows for a cage rearrangement to be described as the system passing through $h=0$, regardless of which particle passes between the others.  Finally, we note that this definition maps all rotationally symmetric states to the same value of $h$.

Shown in Fig.~\ref{fig:traj} are trajectories in $h$-space for three values of $\epsilon$.  As $\epsilon$ increases from top to bottom, the length of time between cage breaking events decreases, as one would expect.  For these and all other trajectories, we find systems spend the majority of time localized around $h = \pm 2$, commensurate with a particle diameter.  The strength of the localization can be inferred from  the fluctuations about $h=\pm 2$, which increase with increasing $\epsilon$. Hence, as the system size increases, the strength of the localization decreases and cage breaking events become more frequent.

\begin{figure}\includegraphics[width=0.48\textwidth]{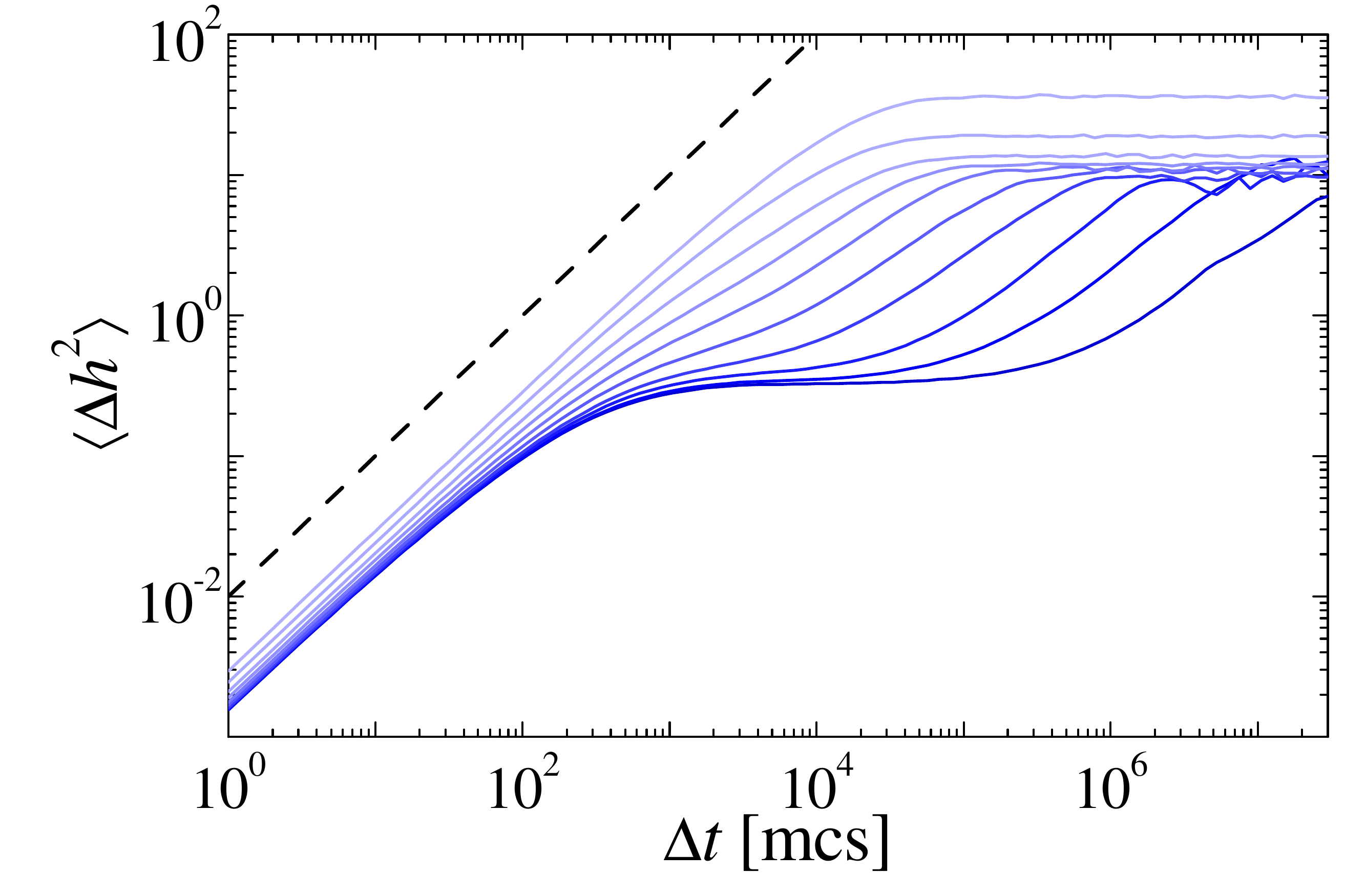}
 \caption{(Color online)  Mean-square displacement in $h$-space for systems with (bottom to top, dark gray to light gray) $\epsilon = 0.045,\ 0.06,\ 0.10,\ 0.15,\ 0.25,\ 0.40,\ 0.61,\ 1.0,\ 2.0,\ \rm{and}\ 4.0$.   Dashed line has a slope of one.} \label{fig:msh}
\end{figure}

To quantify motions in $h$-space, we define a one-dimensional mean-square displacement (MSD), $\langle \Delta h^2(\Delta t) \rangle = \langle \left[h(t+\Delta t) - h(t)\right] ^2 \rangle$, where the average extends over equivalent lag times $\Delta t$.  Shown in Fig.~\ref{fig:msh} are $\langle \Delta h^2 (\Delta t) \rangle$ for various $\epsilon$.  For all system sizes, we observe diffusive behavior on short time scales such that $\langle \Delta h^2 (\Delta t) \rangle \propto \Delta t$, however effects of finite system size are apparent from differences in the intercept at $\Delta t = 1$, as will be discussed shortly.   

For the largest $\epsilon$, $\langle \Delta h^2 (\Delta t) \rangle$ retains diffusive behavior until finally plateauing due to the finite size of the system.  However, the onset of distinctly subdiffusive behavior and development of increasingly long plateaus are apparent as $\epsilon$ is decreased.  For the smallest values of $\epsilon$ in Fig.~\ref{fig:msh}, $\langle \Delta h^2 (\Delta t) \rangle$ is qualitatively similar to the MSDs of supercooled liquids and glass formers, where the plateau indicates timescales over which caging occurs.    The upturn of $\langle \Delta h^2 (\Delta t) \rangle$ indicates that cage breaking eventually occurs for all systems.

\begin{figure}[h]\includegraphics[width=0.48\textwidth]{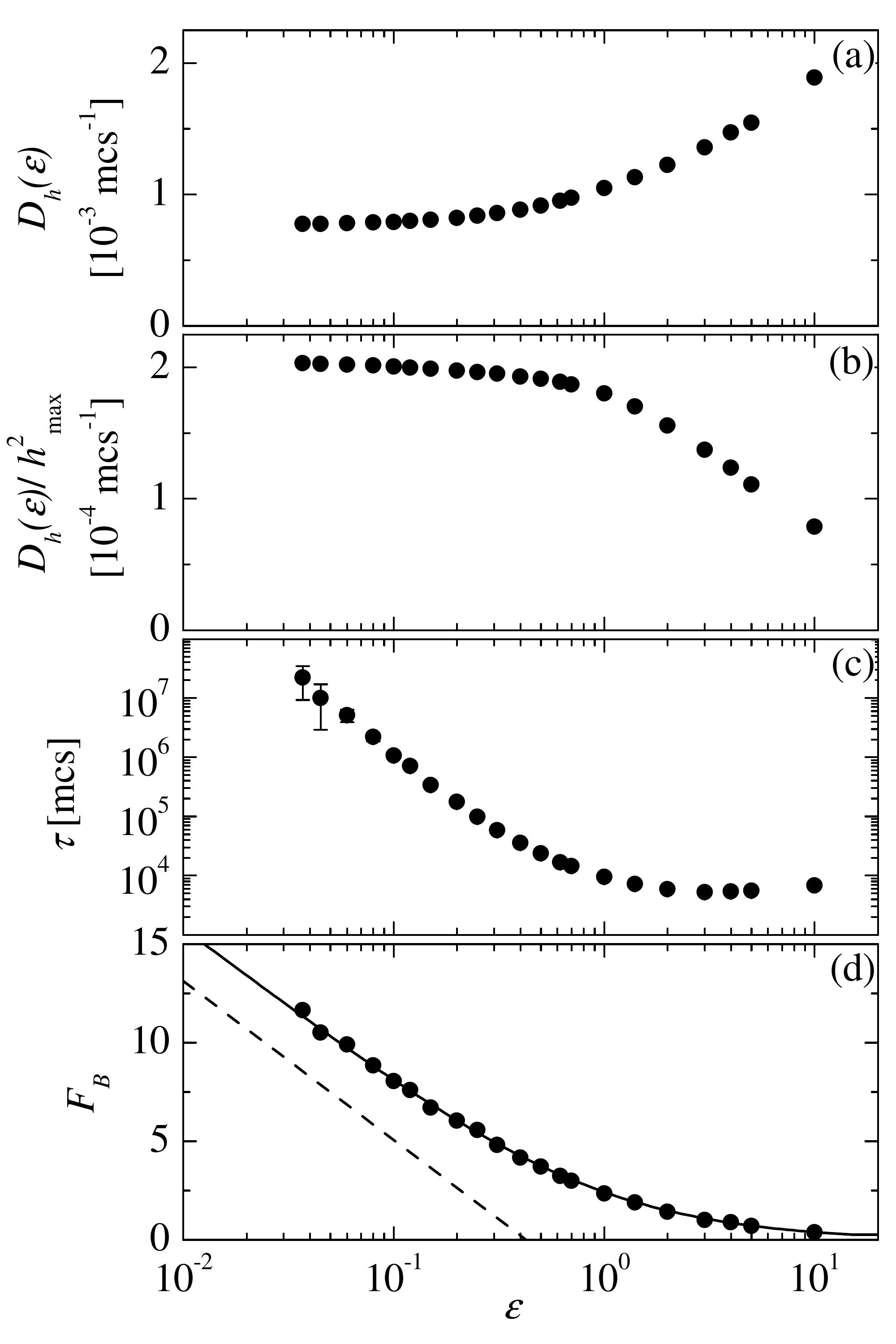}
 \caption{(a) Short time diffusion coefficient scaled by squared particle size. (b) Short time diffusion coefficient scaled by squared system size. (c) Average time between cage breaking events. Error bars are calculated from the statistical uncertainty based on the number of events. Where not visible, error bars are smaller than the symbol.  (d) Height of the free energy barrier as a function of $\epsilon$.  Circles are the results from simulations and the solid line is from the calculations described in Sec.~\ref{sec:enland}.  Dashed line grows as $ \ln{ [\epsilon^{-7/2} ] }$ and shows scaling behavior as $\epsilon \rightarrow 0$.} \label{fig:fourplot}
\end{figure}

Shown in Fig.~\ref{fig:fourplot}(a) are the short time diffusion coefficients in $h$-space, defined as,

\begin{equation}
\label{eqn:shortd}
D_h = \frac{\langle \Delta h^2(\Delta t = 1) \rangle}{ 2}.
\end{equation}
\noindent In Cartesian space, the short time diffusion coefficient is a constant with no spatial or system size dependence. In our 1D coordinate system, however, the diffusion coefficient becomes a function of space (discussed in Sec.~\ref{sec:enland}) and system size.  For small systems, $D_h$ is relatively constant, though we do find a weak linear dependence on $h_{\rm max}$ (not shown).  As $\epsilon$ increases, however, $D_h$ increases without bound, again only as  a consequence of our definition of $h$.  In large systems, two particles may be near each other, while the other is some distance away.  In this case, a small displacement of one of the proximal particles in Cartesian space can translate into a very large displacement in $h$-space.  Hence, the unbounded increase of $D_h$ with $\epsilon$ is not surprising.  If the diffusion coefficient is normalized by the squared system size, however, as in Fig.~\ref{fig:fourplot}(b), we find a trend opposite to that seen in (a).  These values at small $\epsilon$ are again relatively constant, but decrease significantly as the system size increases.  Thus, as $\epsilon$ increases, the system explores absolute space more quickly, but relative to the size of the system, the exploration of space occurs slower.

To quantify a relaxation time scale, or the time the system spends caged, we define a transition time scale $\tau$ as the average time needed to cross $h=0$ from positive $h$ to negative $h$, or vice versa.  As highlighted in Fig.~\ref{fig:traj}, there can be small fluctuations about $h=0$ that are not true cage breaking events.  To minimize biasing $\tau$ toward lower timescales due to this sort of rattling, we stipulate that once $h=0$ is crossed, the system must move a further distance $l^*$ before returning.  Otherwise, no crossing event is registered.  If $F(h)$ defines the free energy landscape (see Sec.~\ref{sec:enland} for more details), the distance $l^*$ is calculated as 

\begin{equation}
\label{eqn:lstar}
l^* = \displaystyle \left(\frac{\int\limits_{-2}^2 h^2 \ F(h) \ dh}{\int\limits_{-2}^2 F(h) \ dh}\right) ^{1/2},
\end{equation}

\noindent that is, the standard deviation of the energy landscape between $[-2,2]$.  As will be discussed in Sec.~\ref{sec:enland}, the height of the energy barrier decreases with system size but extends over exactly this domain for all systems. Hence, Eqn.~(\ref{eqn:lstar}) provides a consistent length scale for where the energy barrier drops to  $\approx 3/5$ of its peak value (further details of calculating $F_B$ are given in Sec.~\ref{sec:enland}).  Shown in Fig.~\ref{fig:fourplot}(c), as $\epsilon$ decreases from 1 toward 0.1, the relaxation time increases, by approximately two orders of magnitude, and continues to grow dramatically as the system becomes smaller.   The smallest relaxation time occurs when $\epsilon \approx 3.0$.  For $\epsilon > 3.0$, motion through the landscape is limited only by diffusion and so the relaxation time increases as a result of the large system size.

\section{Energy Landscape}
\label{sec:enland}

From Eqn.~(\ref{eqn:fts}), determining the energy landscape is only a matter of calculating the entropy for each macrostate $h$.
For simplicity, we set $k_BT \equiv 1$ and write the free energy of a state relative to the ground state as, 
\begin{eqnarray}\label{eqn:fcalc}
F(h) &=& -T\left[S(h) - S_0\right]\nonumber \\
&=&-\ln{\left[n(h)/n_0\right]},
\end{eqnarray}
\noindent where 0 subscripts refer to the ground state at a given $\epsilon$, and $n(h)$ is the number of states that map to the same $h$.  We calculate the number of states or multiplicity $n(h)$ by integrating over the space of allowed configurations $\Omega$ of the three disks while maintaining a fixed $h$.  In general, this can be written

\begin{equation}\label{eqn:nofh}
n(h) = \int\limits_\Omega d\vec{r}_1\  d\vec{r}_2\  d\vec{r}_3\  \delta[h - H(\vec{r}_1,\vec{r}_2,\vec{r}_3)].
\end{equation}

\noindent where the function $H(\vec{r}_1,\vec{r}_2,\vec{r}_3)$ calculates the value of $h$ given the coordinates of the three disks, and $\delta$ is the Dirac delta function.  The expression in Eqn.~(\ref{eqn:nofh}) can be reduced to a 1D integral which we integrate numerically.  We then calculate $F(h)$ using Eqn.~(\ref{eqn:fcalc}).  Further details of calculating $n(h)$ are given in the appendix.

\begin{figure}\includegraphics[width=0.48\textwidth]{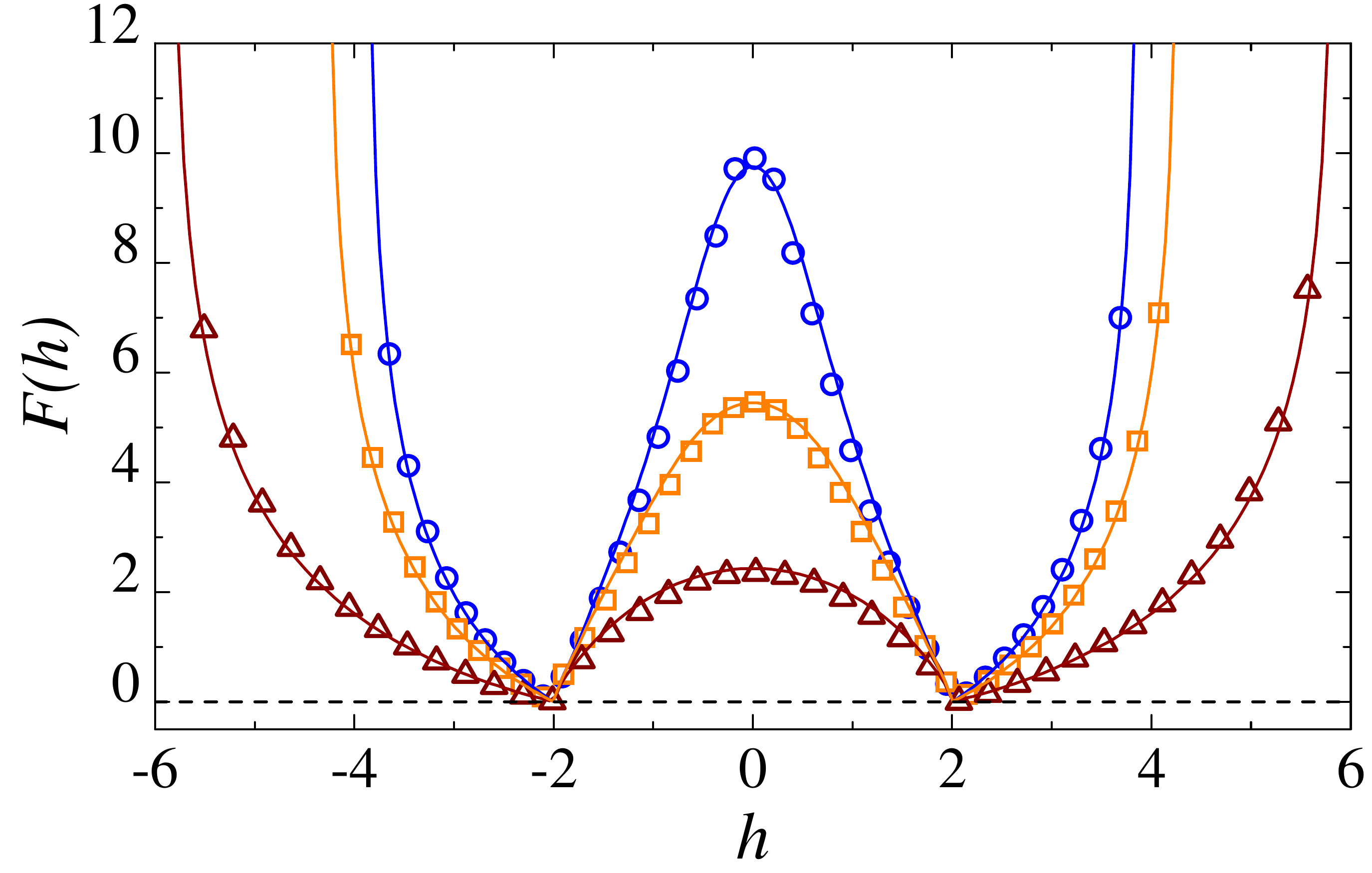}
 \caption{(Color online) Comparison between theoretical energy landscapes (solid lines) and results from simulations (symbols).  Circles: $\epsilon = 0.06$, squares: $\epsilon = 0.25$, triangles: $\epsilon = 1.00$. }\label{fig:eland}
\end{figure}

 To determine the distribution $n(h)$ from simulations, we directly count the number of states by constructing histograms of $h$ with small bins~\cite{threecircfootnote1}.   At this point, $F(h)$ can be computed directly using Eqn~(\ref{eqn:fcalc}).

In Fig.~\ref{fig:eland}, we compare analytical calculations of energy landscapes (solid lines) to those resulting from simulations (symbols).  In all cases, the agreement between theory and simulation is excellent and so we are confident that the number of simulation steps adequately samples the configuration space.  Energy landscapes for all values of $\epsilon$ are symmetric about $h=0$ and are double-welled, with infinitely high barriers at $h=h_{\rm max}$, corresponding to particles being unable to escape the corral.  The height of the energy barrier $F_B$, shown in Fig.~\ref{fig:fourplot}(d), increases as $\epsilon$ decreases and diverges as $\epsilon \rightarrow 0$. As $\epsilon \rightarrow 0$, the height of the barrier grows as $\ln\left[\epsilon ^{-7/2}\right]$, which can be predicted analytically and is described in the appendix.

The most probable $h$ for {\it any} corral size occurs at $h = \pm 2$, as indicated by locations of the minima in Fig.~\ref{fig:eland}.  Inspecting Fig.~\ref{fig:schematic}(b), we see that there are more ways to place disk 3 at a value of $h$ corresponding the cusp of the allowed region than anywhere else.  For $|h| < 2$, the number of states is limited by configurations where particles overlap, whereas this constraint disappears for $|h| \geq 2$.

Given the energy landscape, we can now ascribe the localization about $h = \pm 2$ seen in Fig.~\ref{fig:traj} to the system being trapped in one of two local energy minima.  Furthermore, we see in Fig.~\ref{fig:eland} that the landscape broadens outward as $\epsilon$ increases, which explains the increase in fluctuations shown in Fig.~\ref{fig:traj}.  Inspecting Fig.~\ref{fig:msh}, we see that a longer plateau in $\langle\Delta h^2 \rangle$ equates to a larger free energy barrier, and thus corresponds to the system being constrained at low $\epsilon$ and short times to explore only those regions near a minimum in the energy landscape.  Given enough time, a large thermal fluctuation allows the system to cross the energy barrier, producing the upturn in $\langle \Delta h^2 (\Delta t) \rangle$.

\begin{figure}\includegraphics[width=0.48\textwidth]{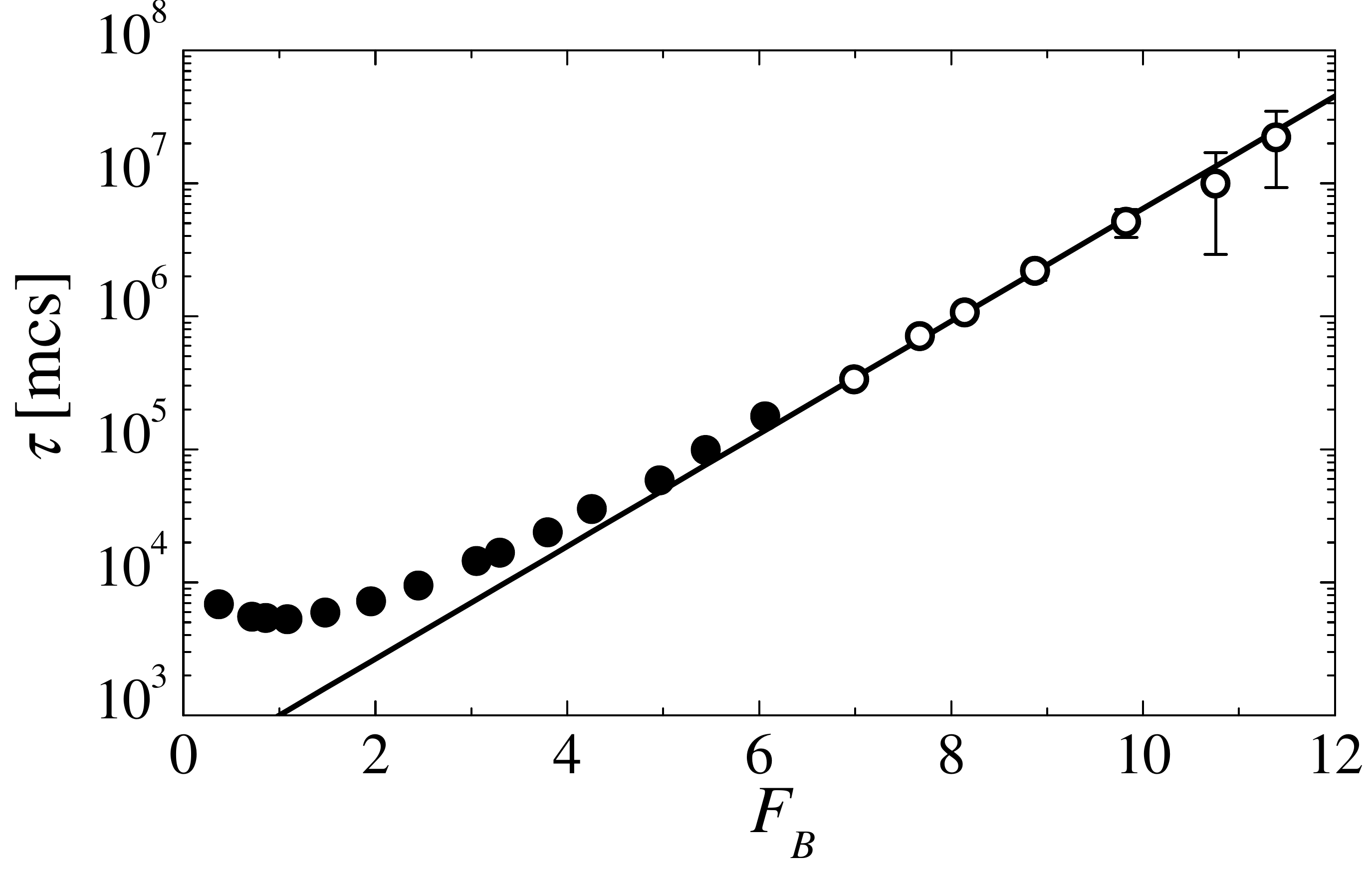}
 \caption{Symbols: Measured relaxation time as a function of theoretical free energy barrier height.  Solid line is $378\exp{(\beta F_B)}$, with $\beta = 0.975 \pm .04$, and comes from a weighted fit to the open symbols.  To within statistical error, transition times grow Arrheniusly with the height of the energy barrier when $F_B \gtrsim 7.0$.} \label{fig:relaxationtimes}
\end{figure}

We are also now in a position to relate the energy landscape to the previously measured relaxation times.  Shown in Fig.~\ref{fig:relaxationtimes}, we find that these relaxation times scale Arrheniusly with barrier height, to within statistical error, when $F_B \gtrsim 7.0$.  As shown in Fig.~\ref{fig:traj}, fluctuations about $|h| = 2$ grow with increasing $\epsilon$.   Additionally, when $\epsilon$ is small, the energy barrier is large.  Hence, for large values of $F_B$, the system finds it difficult to explore large values of $|h|$ and the dominating factor in relaxation is the height of the energy barrier.

In contrast for larger $\epsilon$, the energy barrier is smaller and fluctuations about $|h|= 2$ are much more significant. In these cases, the system finds it easier to wander to larger values of $|h|$ and must first diffuse toward the barrier before crossing.  This accounts for the deviation from Arrhenius behavior toward longer relaxation times at smaller $F_B$.  When $F_B \lesssim 1.0$, relaxation becomes essentially independent of the barrier height and instead depends only on diffusion and system size.

\begin{figure}\includegraphics[width=0.48\textwidth]{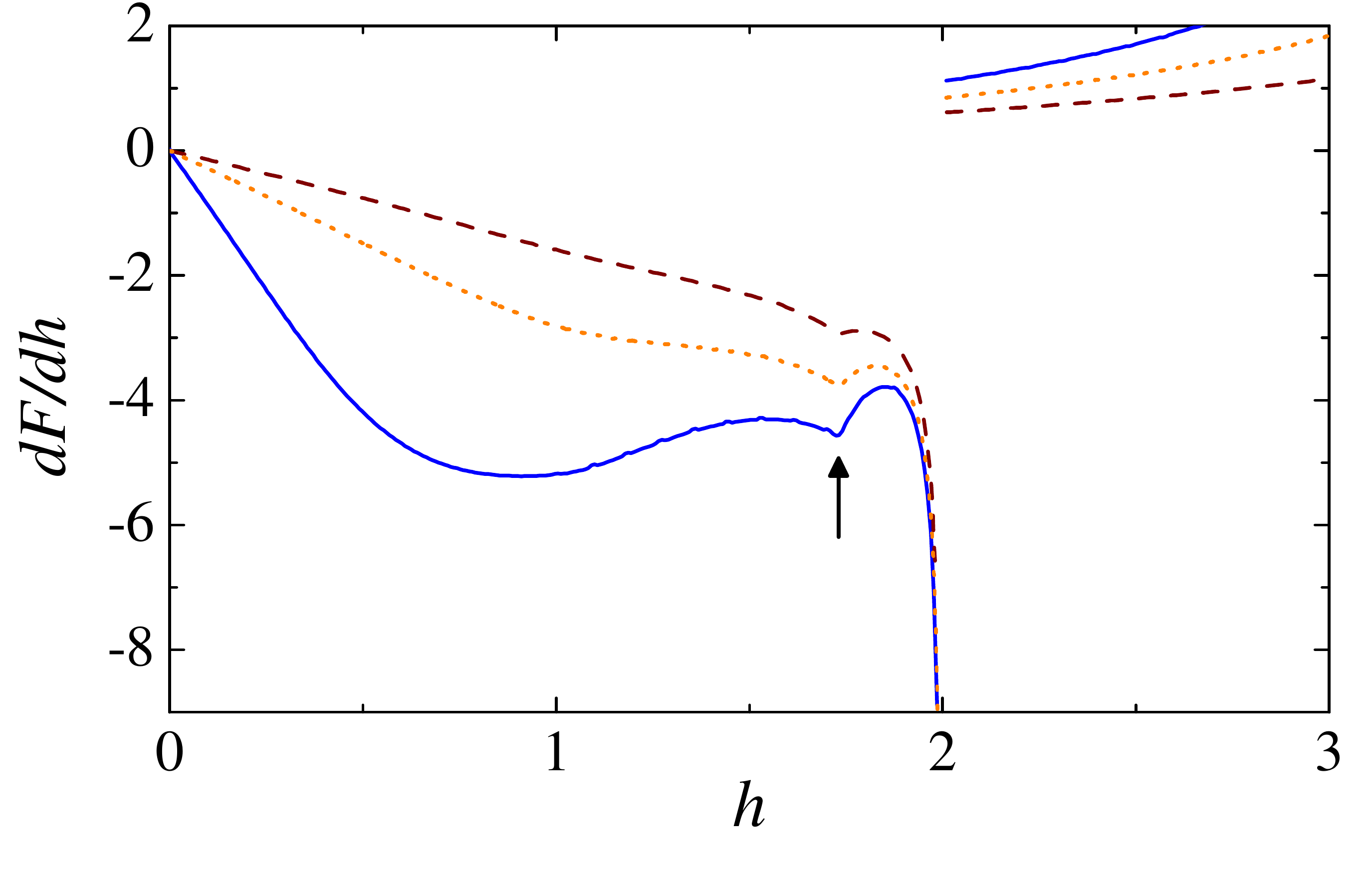}
 \caption{(Color online) First derivatives of $F(h)$ for $\epsilon = $ 0.10 (solid line), 0.31 (dotted line), and 0.62 (dashed line).  Discontinuities exist at $h=\pm 2$.  Additionally, a discontinuity exists in the second derivative at $h = \pm \sqrt{3}$, as indicated by the arrow.}  \label{fig:dfdh}
\end{figure}

In the overdamped limit of Brownian motion, Kramers rate theory~\cite{kramers40} states that the time scale for crossing an energy barrier behaves as~\cite{kramers50, gardinerbook}

\begin{equation}
\label{eqn:kramers}
\tau \propto (D\omega_m \omega_b)^{-1} \exp{(\Delta E / k_BT)}, 
\end{equation}

\noindent where $D$ is the diffusion constant, $\omega_m$ and $\omega_b$ are the curvature at the minimum and barrier, respectively, and $\Delta E \gg k_BT$ is the height of the energy barrier.  In our case, $\omega_m$ is not defined.  The energy landscapes shown in Fig.~\ref{fig:eland} are everywhere continuous, however they are {\it not} everywhere differentiable.   In Fig.~\ref{fig:dfdh}, we show the first derivative of $F(h)$ for $\epsilon = 0.10,$ $0.31$, and $0.62$.  As a consequence of confinement and the projection in $h$-space, a kink in the free energy curve arises for all systems at $h = \pm 2$, and thus the first spatial derivative of free energy is discontinuous. The origins of the discontinuities are discussed in the appendix. Though the curves are steep, we point out that all values of $dF/dh$ in Fig.~\ref{fig:dfdh} are finite.  The discontinuity in $dF/dh$ shows that $\omega_m$ is not defined and so the minimum cannot be approximated as harmonic.   Thus, simple expressions from Kramers rate theory are not able to predict the transition times shown in Fig.~\ref{fig:relaxationtimes}.   Additionally for all systems, we observe a kink in $dF/dh$ at $h=\pm \sqrt{3}$, and therefore the second derivative of the free energy  $d^2F/dh^2$ is discontinuous at these points.

\begin{figure}\includegraphics[width=0.48\textwidth]{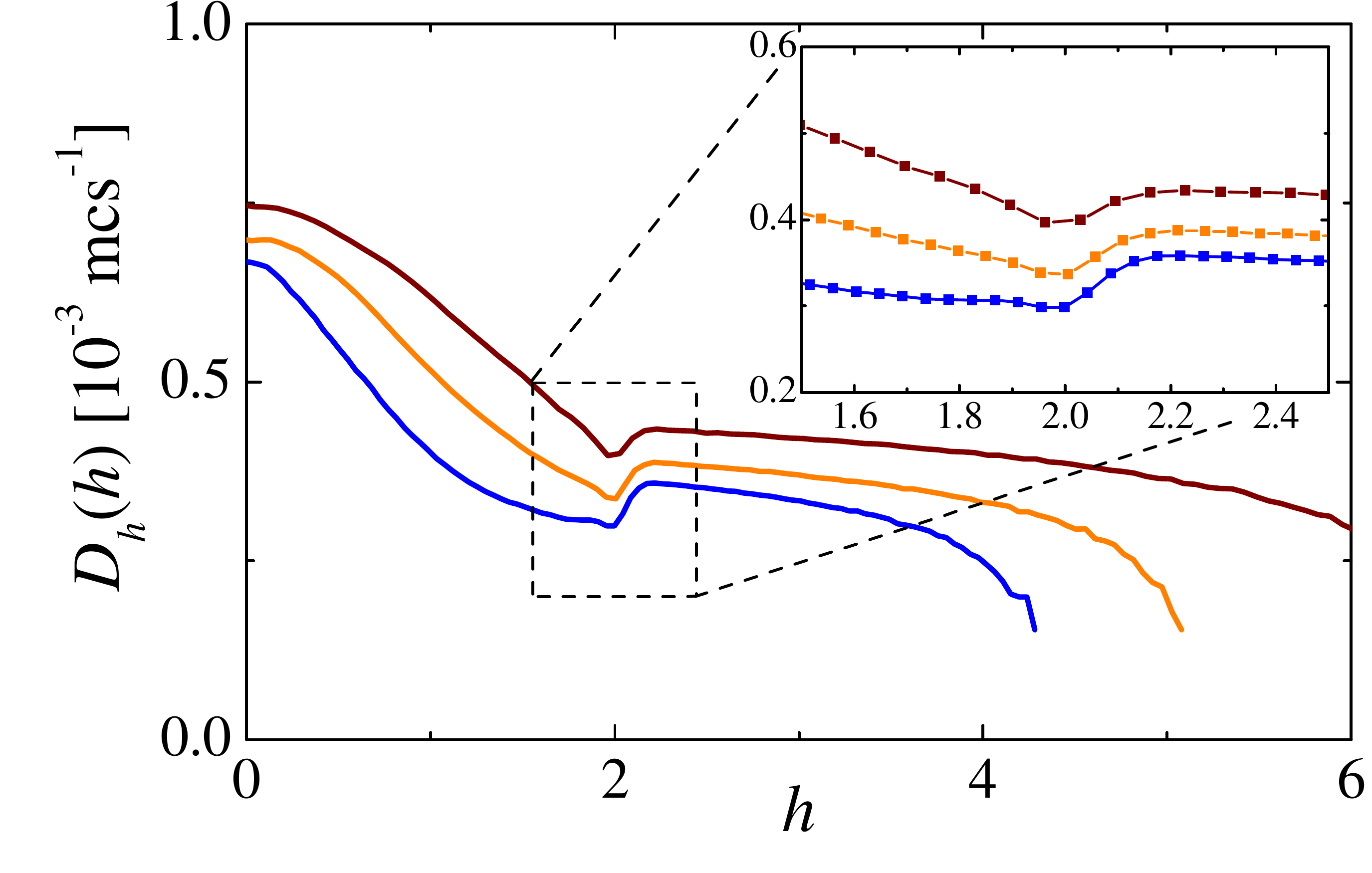}
 \caption{(Color online) Spatially dependent diffusion coefficients for (bottom to top) $\epsilon = 0.31,\ 0.70,\ {\rm and} \ 1.40$ for $h > 0$. Inset: a closer view of region near the kink in $F(h)$.}  \label{fig:dofh}
\end{figure}

As mentioned previously, the diffusion coefficient in $h$-space is spatially dependent.  We calculate a spatial diffusion coefficient, $D_h(h)$ as in Eqn.~(\ref{eqn:shortd}) over a range $\left[h-\delta h,h+\delta h \right]$, where $\delta h$ is $h_{\rm max}/100$.  Near $h=\pm 2$, we measure the limiting value of the variance up to but not including these points.  Shown in Fig.~\ref{fig:dofh} for $\epsilon =0.31,\ 0.70,\ {\rm and}\ 1.40$, the spatially dependent diffusion coefficients appear to be continuous over all permitted values of $h$.  To within our resolution, however, we cannot discern whether or not $D_h(h)$ is everywhere differentiable.  

\begin{figure}\includegraphics[width=0.48\textwidth]{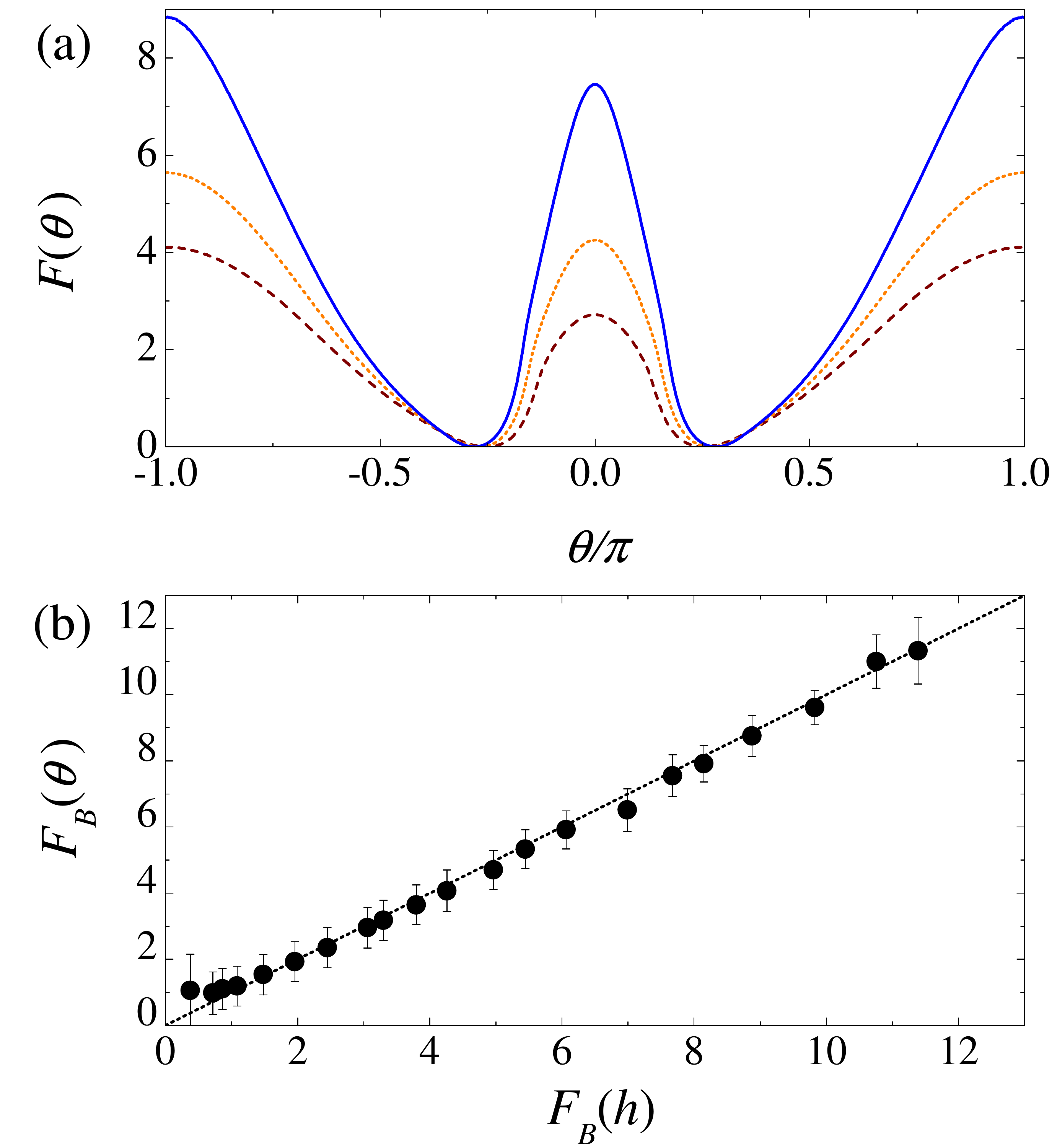}
 \caption{(Color online) (a) Free energy landscape in terms of $\theta$ for $\epsilon$ = 0.10 (solid line), 0.31 (dotted \ line), and 0.62 (dashed \ line). (b) Heights of energy barriers for all $\epsilon$ in $\theta$-space versus the barrier height in $h$-space, calculated as described in the text.  Error bars indicate the uncertainty in local quadratic fits of $n(0)$ and $n_0$.  Dotted line is $F_B(\theta) = F_B(h)$.}  \label{fig:tenvhen}
\end{figure}

Our focus thus far has used $h$ as a means of capturing the relevant cage breaking dynamics, but of course, one can imagine many options when reducing the configuration space from six dimensions down to one.  For consistency, it is worth checking that the results shown above are not dependent on the choice of variable.  We thus define a second variable $\theta$, shown in Fig.~\ref{fig:schematic}, that describes the angle between vectors $\vec{r}_{21} = \vec{r}_2-\vec{r}_1$ and $\vec{r}_{31} = \vec{r}_3 - \vec{r}_1$.  A cage breaking event can then be defined when the system crosses $\theta = 0$ or $\theta = \pm \pi$.  Free energy landscapes in $\theta$-space are shown in Fig.~\ref{fig:tenvhen}(a)  for $\epsilon = 0.10, \ 0.31,$ and $ 0.62$~\cite{threecircfootnote2}.  We find the energy landscape is again double-welled, reflecting the two possible caged states, and has even symmetry about $\theta = 0$ between $[-\pi,\pi]$.  Here, unlike $F(h)$, we find that the locations of the minima in the energy landscape are not constant but move to smaller values of $\theta$ as $\epsilon$ increases.  For example in Fig.~\ref{fig:tenvhen}(a), minima shift from $\theta = \pm .245\pi$ at $\epsilon = 0.10$ to $\theta = \pm 0.225\pi$ at $\epsilon = 0.62$. 

As shown in Fig.~\ref{fig:tenvhen}(a), there are two energy barriers in $\theta$-space as opposed to the the single barrier in $h$-space.  While the representation has changed from variable $h$ to $\theta$, the fundamental problem of caging has not, i.e. cage breaking is equally difficult or equally easy independent of the representation.  Hence, for these measurements of energy to be consistent, there must be a relationship between the energy barriers measured in the two different coordinate systems.

In Fig.~\ref{fig:tenvhen}(b), we show the height of the free energy barriers $\theta$-space $\epsilon$ as a function of the previously measured barrier heights in $h$-space.  The heights of the barriers are determined from the same simulation data that was used to calculate $F(h)$.  To calculate $F_B(\theta)$, we consider the average probability that the system is poised to cage break, either at $\theta = 0$ or $\theta = \pi$.  This yields 

\begin{equation}\label{eqn:fbtheta}
F_B(\theta) = -\log{ \{[n(0) + n(\pi) ]/[ 2 n_0 ]\} }.
\end{equation}

\noindent As shown in the appendix, $n(0) = 4n(\pi)$.  Using this fact, we only measure $n(0)$ and $n_0$ to compute $F_B(\theta)$ in Eqn.~(\ref{eqn:fbtheta}).  The heights of the energy barriers in $\theta$-space shown in Fig.~\ref{fig:tenvhen}(b) are in excellent agreement with those using $h$ as a coordinate.  The consistency between these two measurements demonstrate that both coodinates adequately describe caging and that the measured free energies are indeed those relevant to cage breaking in our system.

\begin{figure}\includegraphics[width=0.48\textwidth]{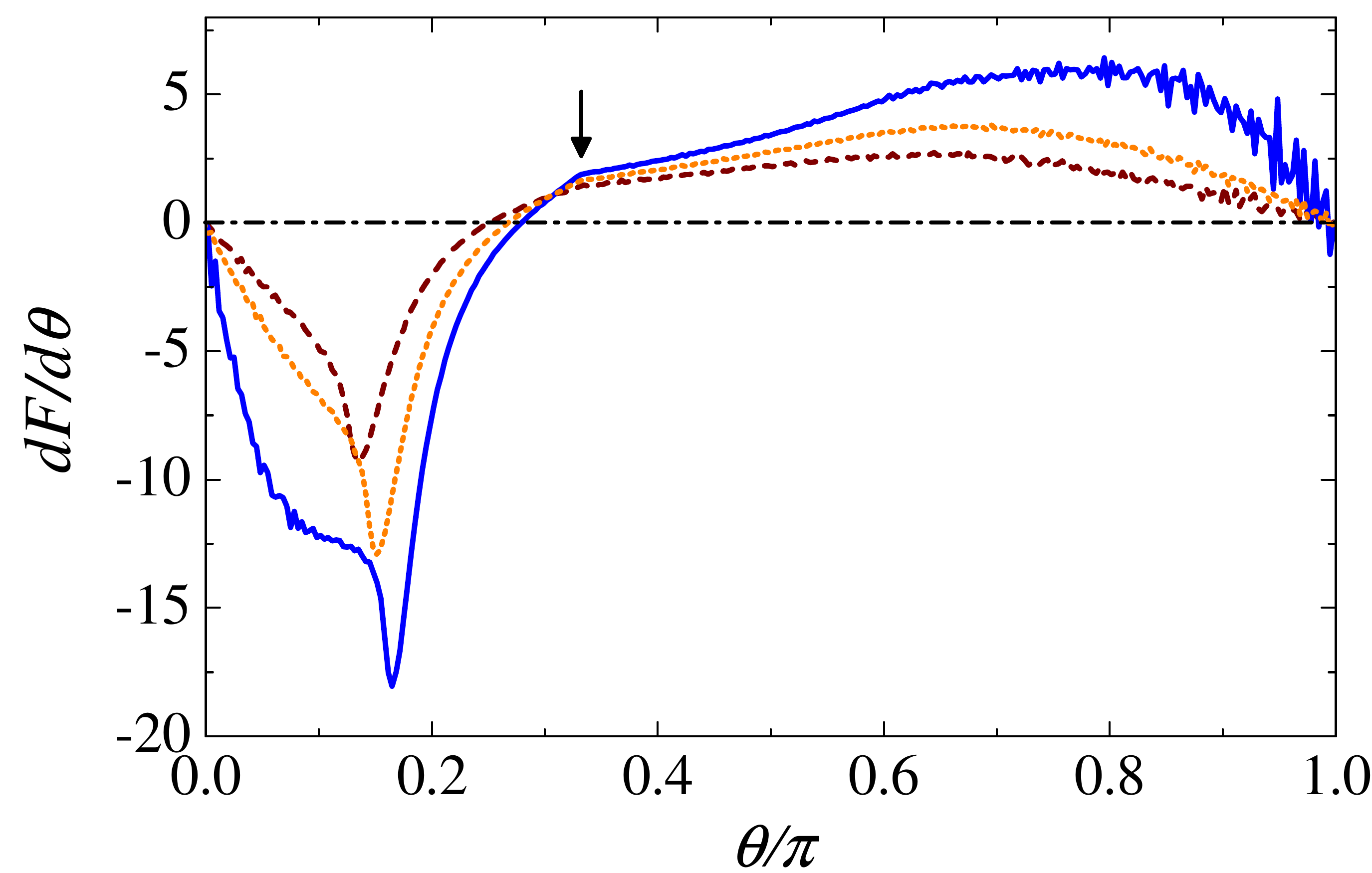}
 \caption{(Color online) First derivative of $F(\theta)$ for $\epsilon$ = 0.10 (solid line), 0.31 (dotted line), and 0.62 (dashed line).  All first derivatives exhibit a subtle kink at $\theta = \pm \pi/3$ indicated by the arrow. Horizontal dashed-dotted line is at $dF/d\theta$ = 0, and helps illustrate the gradual movement of the minimum to smaller values of $\theta$ for increasing $\epsilon$.}  \label{fig:deltheta}
\end{figure}

 As in the case of $F(h)$, $F(\theta)$ is everywhere continuous.  We show in Fig.~\ref{fig:deltheta} the first derivative of $F(\theta)$.  Here, curves are obtained by numerically differentiating those in Fig.~\ref{fig:tenvhen}(a). In contrast to $F(h)$, $dF/d\theta$ is also everywhere continuous, and so $F(\theta)$ is also everywhere differentiable.  Without an analytical calculation of $F(\theta)$ we cannot draw definitive conclusions about the continuity of $d^2F/d\theta^2$.  However, we do find for each $\epsilon$ the hint of a subtle kink in $dF/d\theta$ located at $\theta = \pm \pi/3$, as well as a second $\epsilon$-dependent region where the slope of $dF/d\theta$ changes sharply, e.g. at approximately $\theta = 0.16 \pi$ for $\epsilon = 0.10$ in Fig.~\ref{fig:deltheta}. As in $h$-space, diffusion in $\theta$-space also exhibits a complicated spatial dependence (not shown). One assumption leading to Eqn.~(\ref{eqn:kramers}) is that $D$ is spatially independent.  Predictions for $\tau$ which incorporate spatially dependent diffusion are highly non-trivial~\cite{gardinerbook}.

\section{Discussion and Conclusions}

We have introduced a  model system composed of three hard disks confined to a circular corral.  Though simple, this system exhibits caging and cage breaking behaviors, reminiscent of dense liquids, and allows us to exactly calculate a free energy landscape in terms of a single system coordinate.  Respectively, caging and cage breaking can then be understood as the system becoming trapped in local energy minima and eventually being thermally excited over an energy barrier.

 As the size of the system decreases, exploration of the configuration space is increasingly hindered by disk-wall and disk-disk interactions, though cage rearrangements do eventually occur for all $\epsilon > 0$.   Arrhenius scaling describes the transition time $\tau$ between wells when the energy barrier is large and the system is small.  However, this scaling fails for smaller energy barriers, when diffusion becomes the limiting factor in relaxation.

The coordinate(s) one uses to express the energy landscape must capture the behavior one is studying~\cite{frenkel92,frenkel96,frenkel97,heuer97,walesbook,krivov04,altis08,liwo09,bevan11}.  Those used in the text successfully describe motion from one cage to the other, however, there are any number of coordinates one could use that bear no relevance to caging.  For example, a calculation of the collective radius of gyration of the disks may yield the same result for configurations near cage breaking as for those far away.  In the same respect, a particular value of $h$ tells one very little about the actual configuration or spatial location of the disks.  Furthermore, that caging can be described successfully in $h$-space or $\theta$-space demonstrates that projections of the landscape to a lower dimension are not necessarily unique.

Depending on the choice of coordinates, the energy landscape of hard disk and hard sphere systems may be locally non-differentiable.  This observation highlights the notion that projection of a many dimensional landscape down to a lower dimensional representation may introduce artifacts~\cite{liwo09,altis08,krivov04}.  For systems with $N > 3$, it is probable that discontinuities in spatial derivatives of free energy disappear completely, or perhaps are only present for higher order derivatives.  

For {\it any} purely soft core potential, the observed discontinuities will disappear.  For a potential more representative of colloidal suspensions, such as a hard core potential with short ranged repulsion \cite{bryanthardsphere,germain04}, we expect the essence of our results to be valid, but in need of some modification.  For example, a short range repulsion characterized by a length $\delta$ will not significantly affect the free energy when average particle separations are larger than $\delta$.  In these cases, the free energy will be governed essentially by entropy and our results should apply.  For more confined systems, i.e. when particle separations are less than or comparable to $\delta$, whether or not the free energy is dominated by entropy or the potential will depend entirely on the details of the potential.  We are currently investigating these scenarios.

\section{Acknowledgements}
We thank James Kindt for bringing \cite{speedytwodiscs, speedy1999discs} to our attention, and C. Trent Brunson, Stefan Boettcher, and Ilya Nemenman for helpful discussions.  This work was supported by the National Science Foundation under Grant No. CHE-0910707.

\input{appendix}

\bibliography{threecirc5}

\end{document}

%% file: appendix.tex
\section{Appendix}

\begin{figure}[here]
\centerline{
\includegraphics[width=.35\textwidth]{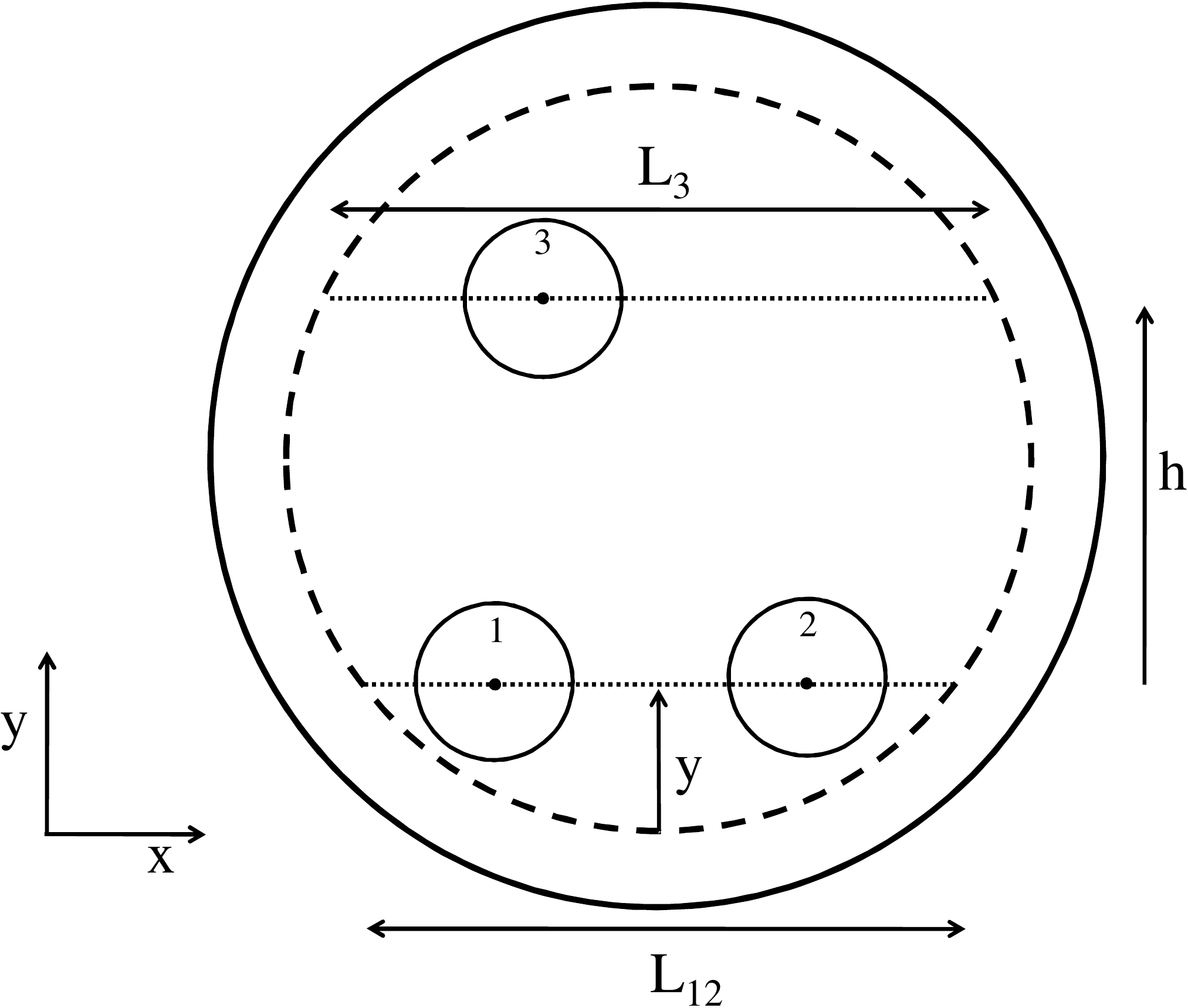}}
\caption{The radius of the solid outer circle is $R_C = R+r$ and is the physical boundary, but the centers of the confined disks can only exist a distance $\leq R$ from the center, indicated by the dashed inner circle.}
\label{fig:caseA}

\end{figure}

Here we present a derivation of $n(h)$ based on geometric arguments.  Because the distribution $n(h)$ has even symmetry about $h = 0$, we directly treat only the cases where $h \ge 0$.  
All three disks are distinguishable and have the same diameter $d$.  In simulations, we used $d=2$, but for the sake of generality here we assume no specific disk size.  The true radius of the corral is $R_C = 3r + \epsilon$, but for simplicity, we define $R = d + \epsilon$ as the radius of the corral accessible to the centers of the disks, as illustrated in Fig.~\ref{fig:caseA}.

While Eqn.~(\ref{eqn:nofh}) is general and exact, it can be further simplified for our system.  Recalling that $H(\vec{r}_1,\vec{r}_2,\vec{r}_3)$ in Eqn.~(\ref{eqn:nofh}) maps all rotationally equivalent states to the same $h$, we may consider only a single orientation of the system, such as the one in Fig.~\ref{fig:caseA}.  We constrain disks 1 and 2 to always lie along the same horizontal line, $L_{12}$, at a distance $y$ above the bottom of the corral.  For a given $h$, disk 3 may lie anywhere along a second horizontal line, $L_3$, at a distance $y+h$ above the bottom of the corral. Thus, the number of states $n(h)$ can be written as the product of integrals

\begin{equation}
\displaystyle n(h) \propto \int dy  \int dx_1 \int dx_2 \int dx_3,
\label{eqn:general}
\end{equation}

\noindent and the problem becomes determining the appropriate limits of integration. For a circle centered at $(0,R)$, we determine the lengths of chords $L_{12}$ and $L_{3}$ to be 

\begin{eqnarray}
L_{12} &=& 2\sqrt{2yR - y^2} \label{eqn:l12} \\
L_3 &=& 2\sqrt{2(y+h)R - (y+h)^2}. \label{eqn:l3}
\end{eqnarray}

\noindent  There are four cases that must be considered to calculate $n(h)$ correctly, each of which is geometrically distinct.

\subsection{Case A  ($h \ge d$)}
\label{sec:caseA}

The first of four cases that must be considered is shown in Fig. \ref{fig:caseA}. This is the simplest case in that disk 3 never comes into contact with disks 1 or 2.  Taking that the chord $L_{12}$ begins at $x = 0$,  the lower and upper bounds of $x_2$ are, respectively $x_1 + d$ and $L_{12}$.  The lower and upper bounds of $x_1$ are $0$ and $L_{12}-d$.  Disk 3 may exist anywhere along the chord $L_{3}$ independently of $x_1$ and $x_2$, so the lower and upper bounds for $x_3$ are $0$ and $L_{3}$.  Therefore,

\begin{eqnarray}
 \displaystyle n_A &\propto& \int\limits_{y_{\rm min}}^{y_{\rm max}}dy  \int\limits_0^{L_3} dx_3 \int\limits_0^{L_{12} - d} dx_1 \int\limits_{x_1 + d}^{L_{12}}dx_2      \nonumber \\
      \displaystyle &=& \int\limits_{y_{\rm min}}^{y_{\rm max}}dy \ L_3 \left[ \frac{(L_{12}-d)^2}{2} \right].
\end{eqnarray}

The lower limit $ y_{\rm min}$ corresponds to when disks 1 and 2 are in contact at the bottom of the corral.  Geometry yields that this occurs at

\begin{equation}
\label{eqn:yminA}
\displaystyle y_{\rm min} = R - \sqrt{R^2 - \frac{d ^2}{4}}
\end{equation}

In this case, the upper limit $y_{\rm max}$ can correspond to $L_{12} = d$ or $L_3 = 0$.  The correct value is the one that minimizes $y_{\rm max}$ and therefore keeps all particle centers within the allowed region.  Solving Eqns.~(\ref{eqn:l12}) and (\ref{eqn:l3}) with these conditions yields, respectively, 

\begin{equation}
\label{eqn:ymaxA}
\displaystyle y_{\rm max} = \min\{2R-h, R + \sqrt{R^2 - \frac{d ^2}{4}}\}
\end{equation}


\subsection{Case B  $\left(h<d , x_3 < x_1 < x_2 \right)$}
\label{sec:caseB}

\begin{figure}[here]
\centerline{
\includegraphics[width=.35\textwidth]{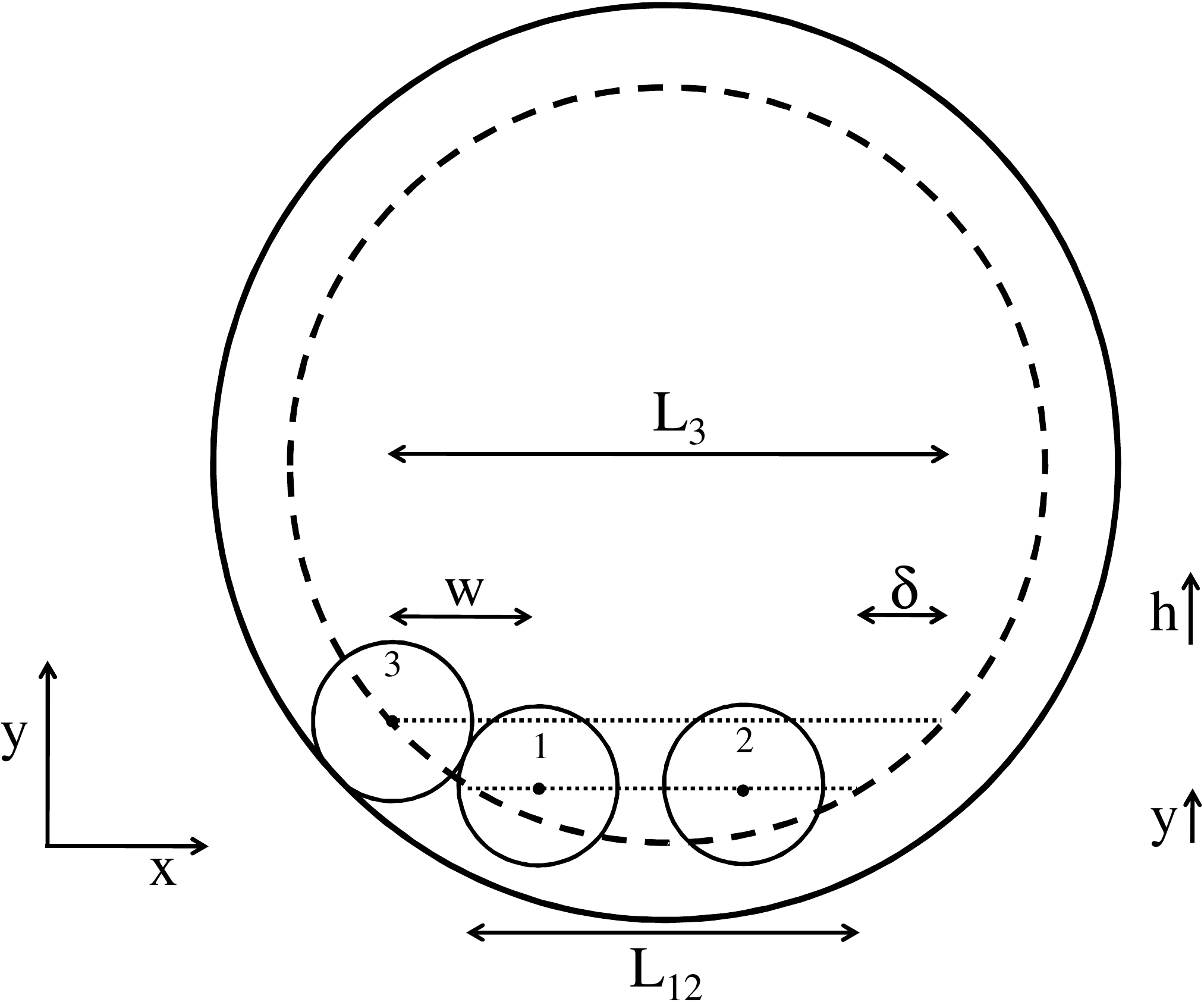}}
\caption{Case B.}
\label{fig:caseB}

\end{figure} 

When $h < d$, disk 3 may come into contact with either disk 1 or disk 2.  In these cases, $n(h)$ must be split into parts and computed in a slightly different manner.  

Three scenarios contribute to $n(h)$: $(x_3 < x_1 < x_2)$ ; $( x_1 < x_3 < x_2)$ ; $(x_1 < x_2 < x_3)$.  The first and last of these scenarios are symmetric and contribute equally to $n(h)$.  The second scenario is presented in sections \ref{sec:caseC} and \ref{sec:caseD}.

For now, we consider only cases where the $x$-coordinate of disk 3 is not between those of disks 1 and 2.  Such a case is shown in Fig. \ref{fig:caseB}.  As stated, there are equally as many states where $x_3 < x_1< x_2$ as there are for $x_1< x_2 < x_3$.  Therefore, we will calculate the multiplicity for only  the states where $x_3 < x_1< x_2$, and finally multiply by two to obtain the contribution to $n(h)$.  We define $w$  as the minimum horizontal separation between disks 1 and 3, given by

\begin{equation}
\label{eqn:defw}
w = \sqrt{d ^2 - h^2}
\end{equation}

\noindent The minimum horizontal separation between disks 1 and 2 remains $d$.  

We also define a quantity $\delta = \left(L_3 - L_{12}\right)/2$, as shown in Fig. \ref{fig:caseB}.  If we take the starting point of chord $L_{12}$ to be 0, then the lower and upper bounds on $x_3$ are, respectively, $\left[ -\delta, L_{12} - d - w \right]$.  From Fig. \ref{fig:caseB}, the limits of integration for $x_2$ are found to be $\left[x_1+d,L_{12} \right]$ and those of $x_1$ are $\left[ x_3+w,L_{12}-d \right]$.  Including the factor of $2$ from the symmetric states, Eqn.~(\ref{eqn:general}) becomes

\begin{eqnarray}
\displaystyle n_B &\propto& 2 \int\limits_{y_{\rm min}}^{y_{\rm max}} dy \int\limits_{(L_{12}-L_3)/2}^{L_{12}-d - w} dx_3 \int\limits_{x_3 + w}^{L_{12}-d}
dx_1 \int\limits_{x_1 + d}^{L_{12}} dx_2 \nonumber \\
\displaystyle &=& 2 \int\limits_{y_{\rm min}}^{y_{\rm max}} dy \ \left(\frac{1}{6}\right) \left( \frac{L_{12}+L_3}{2}-d - w\right)^3
\end{eqnarray}

The limits of integration for $y$ in this case are found from solving the equation 
\begin{equation}
\label{eqn:ylimits}
L_3 = w + d + \delta
\end{equation}

\noindent which after substitution of the various terms yields a quadratic equation in $y$. The limits of integration for $y$ are the roots of this equation, such that the minimum root is $y_{\rm min}$ and the maximum root is $y_{\rm max}$.

\subsection{Case C  $\left( \displaystyle \frac{\sqrt{3}d }{2} < h < d  , x_1 < x_3 < x_2 \right) $ }
\label{sec:caseC}

\begin{figure}[here]
\centerline{
\includegraphics[width=.35\textwidth]{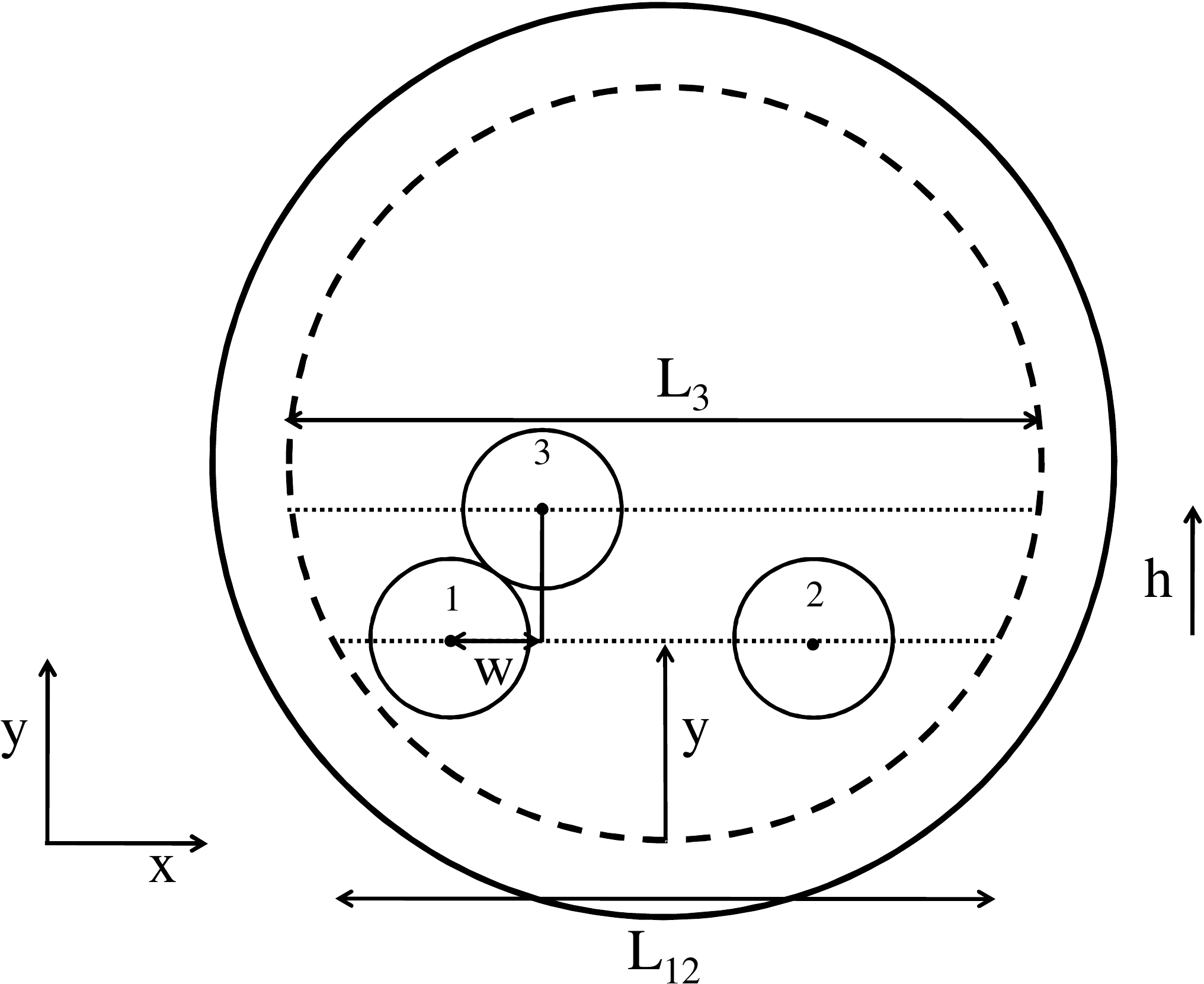}}
\caption{Case C.}
\label{fig:caseC}

\end{figure} 

Complementary to Case B, we now consider the case where $(x_1 < x_3 < x_2)$ and $ \sqrt{3} d /2 < h < d $.  As shown in Fig. \ref{fig:caseC}, the minimum horizontal separation between disk 3 and one of the others is again $w$.  Also, $h$ is large enough that disks 1 and 2 may still contact, hence the minimum separation between disks 1 and 2 is $d$.

Proceeding as before, the limits of integration for $x_1$, $x_2$ and $x_3$ respectively are $\left[ 0,L_{12}-d \right]$,  $\left[x_1+d,L_{12} \right]$, and $\left[x_1+w,x_2-w\right]$.  Eqn. (\ref{eqn:general}) becomes

\begin{eqnarray}
\displaystyle n_C &\propto& \int\limits_{y_{\rm min}}^{y_{\rm max}} dy  \int\limits_{0}^{L_{12}-d} dx_1 \int\limits_{x_1 + d}^{L_{12}} dx_2 \int\limits_{x_1 + w}^{x_2 - w} dx_3 \nonumber \\
\displaystyle &=& \int\limits_{y_{\rm min}}^{y_{\rm max}}dy\ (L_{12} - d)^2 \left[ \frac{(L_{12}-d)}{6} -\left(w - \frac{d}{2} \right)\right] \nonumber \\
\end{eqnarray}

The lower and upper limits of $y$ are found in the same way as in case A.

\subsection{Case D  $\left(0 \leq h \leq \sqrt{3}d/2  , x_1 < x_3 < x_2 \right)$}

\label{sec:caseD}

\begin{figure}[here]
\centerline{
\includegraphics[width=0.35\textwidth]{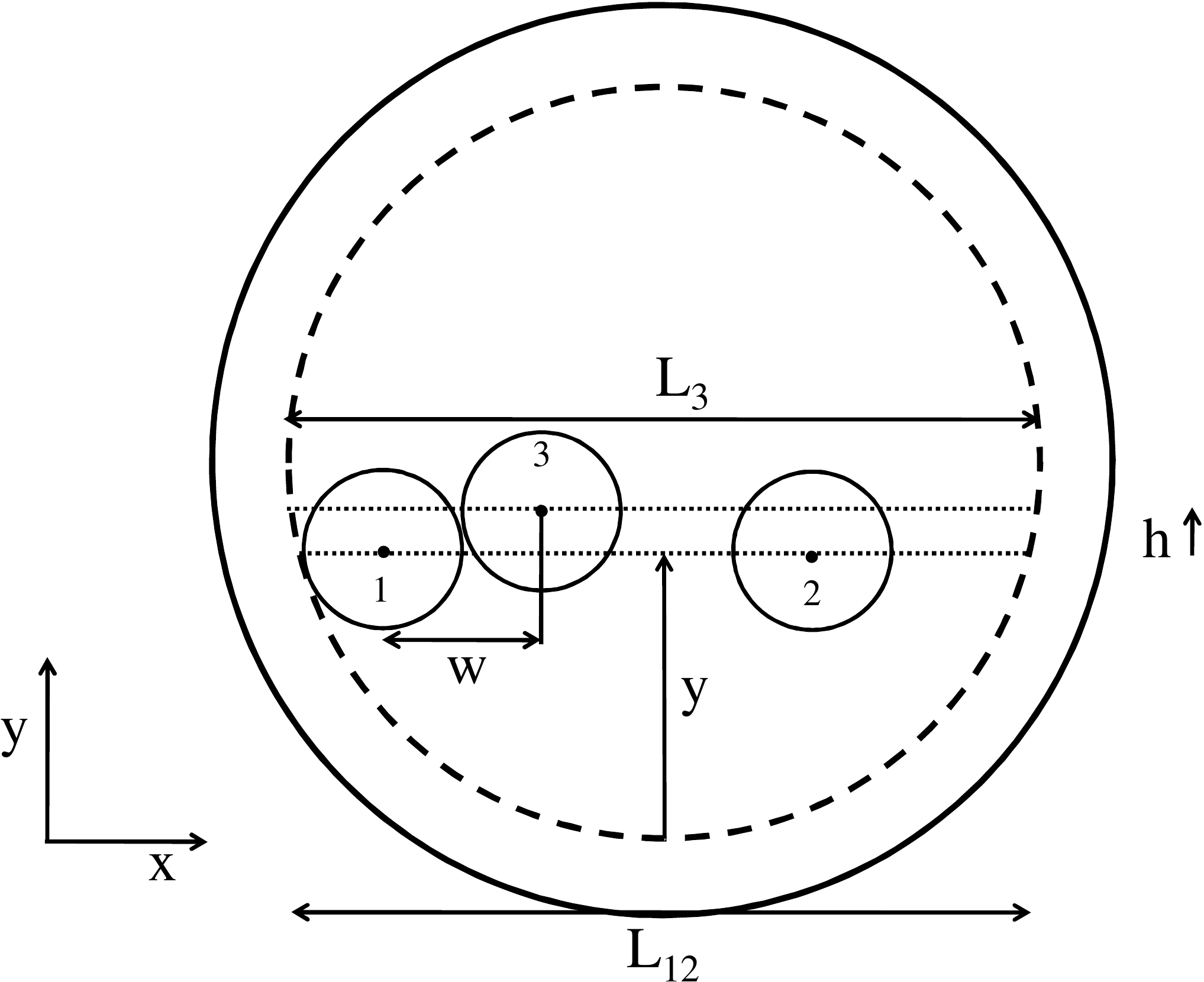}}
\caption{Case D.}
\label{fig:caseD}
\end{figure}

The final case to be considered is also complementary to case B. In Fig. \ref{fig:caseD}, disk 3 is again kept between 1 and 2, only now $h$ is such that disks 1 and 2 never come into contact.  The horizontal separation between 1 and 2 is always $\geq 2w$.  Inspection again yields the limits for $x_1$, $x_2$, and $x_3$ which are $\left[ 0,L_{12}-2w \right]$, $\left[x_3+w,L_{12} \right]$, and $\left[ x_1+w, L_{12}-w \right]$.  Therefore, Eqn.~(\ref{eqn:general}) becomes

\begin{eqnarray}\label{eqn:nofhD}
\displaystyle n_D &\propto& \int\limits_{y_{\rm min}}^{y_{\rm max}} dy \int\limits_{0}^{L_{12}-2w} dx_1 \int\limits_{x_1+w}^{L_{12}-w} dx_3 \int\limits_{x_3 + w}^{L_{12}} dx_2 \nonumber \\
&=&\int\limits_{y_{\rm min}}^{y_{\rm max}} dy \ \left(\frac{1}{6} \right) \left(L_{12}-2w\right)^3
\end{eqnarray}

\noindent The lower limit $y_{\rm min}$ comes from geometry,

\begin{equation}
\label{eqn:yminD}
y_{\rm min} = R - \sqrt{R^2 - w^2},
\end{equation}

\noindent while, similar to case A, the upper limit $y_{\rm max}$ is the minimum value corresponding to either $L_3 = 0$ or the maximum root of $L_{12}= 2w$,

\begin{equation}
\label{eqn:ymaxD}
y_{\rm max} = \min\{2R-h,R+\sqrt{R^2 - w^2}\}.
\end{equation}

Given the results of all four cases, the generalized method to compute $n(h)$ is given by

\begin{equation}\label{eqn:computeh}
n(h) = \left\{ \begin{array}{cl}
  n_A, &\mbox{$h \ge d$} \\
  n_B + n_C, &\mbox{$ \sqrt{3}d/2 < h < d$}\\
  n_B + n_D, &\mbox{$0 \le h \le  \sqrt{3}d/2$}
       \end{array} \right.
\end{equation}

In general, the integrand for each case above is a cumbersome function of $h$, thus we calculate $n(h)$ using numerical integration.  From Eqn.~(\ref{eqn:computeh}), the origin of the kinks in $F(h)$ become more apparent as transitions from one solution regime to another.  For $d=2$, transition points occur at $h = \pm \sqrt{3}$ and  $h = \pm 2$, as described in the text.

\subsection{Behavior of $F_B(h)$ as $\epsilon \rightarrow 0$}

To understand the growth of the energy barrier as $\epsilon \rightarrow 0$, we only need the result of Case D.  Setting $h = 0$ gives $w = d$, and in the limit of small $\epsilon$, transitions occur along the diameter of the corral, which implies $L_{12} \approx 2d + 2\epsilon$.  Eqn. (\ref{eqn:nofhD}) then becomes

\begin{equation}
n(h) \propto \int\limits_{y_{\rm min}}^{y_{\rm max}} dy \ \left(\frac{4}{3} \right) \epsilon^3 \propto \epsilon^3  \Delta y
\end{equation}

where
\begin{equation}
\Delta y = 2\sqrt{R^2 - d^2} = 2\sqrt{2d \epsilon + \epsilon^2}.
\label{barheight3}
\end{equation}

In the limit $\epsilon \rightarrow 0$, $\Delta y$ grows as $\sqrt{\epsilon}$ and so $n(h) \propto \epsilon^{7/2}$.  Therefore, the barrier height grows as

\begin{equation}
F_B \propto -\log{n(h)} \propto -\frac{7}{2} \ln{\epsilon}
\label{barheight4}
\end{equation}

\noindent as confirmed by the data shown in Fig.~\ref{fig:fourplot}(d).


\subsection{Barrier heights in $\theta$}
\begin{figure}[h]
\centerline{
\includegraphics[width=0.4\textwidth]{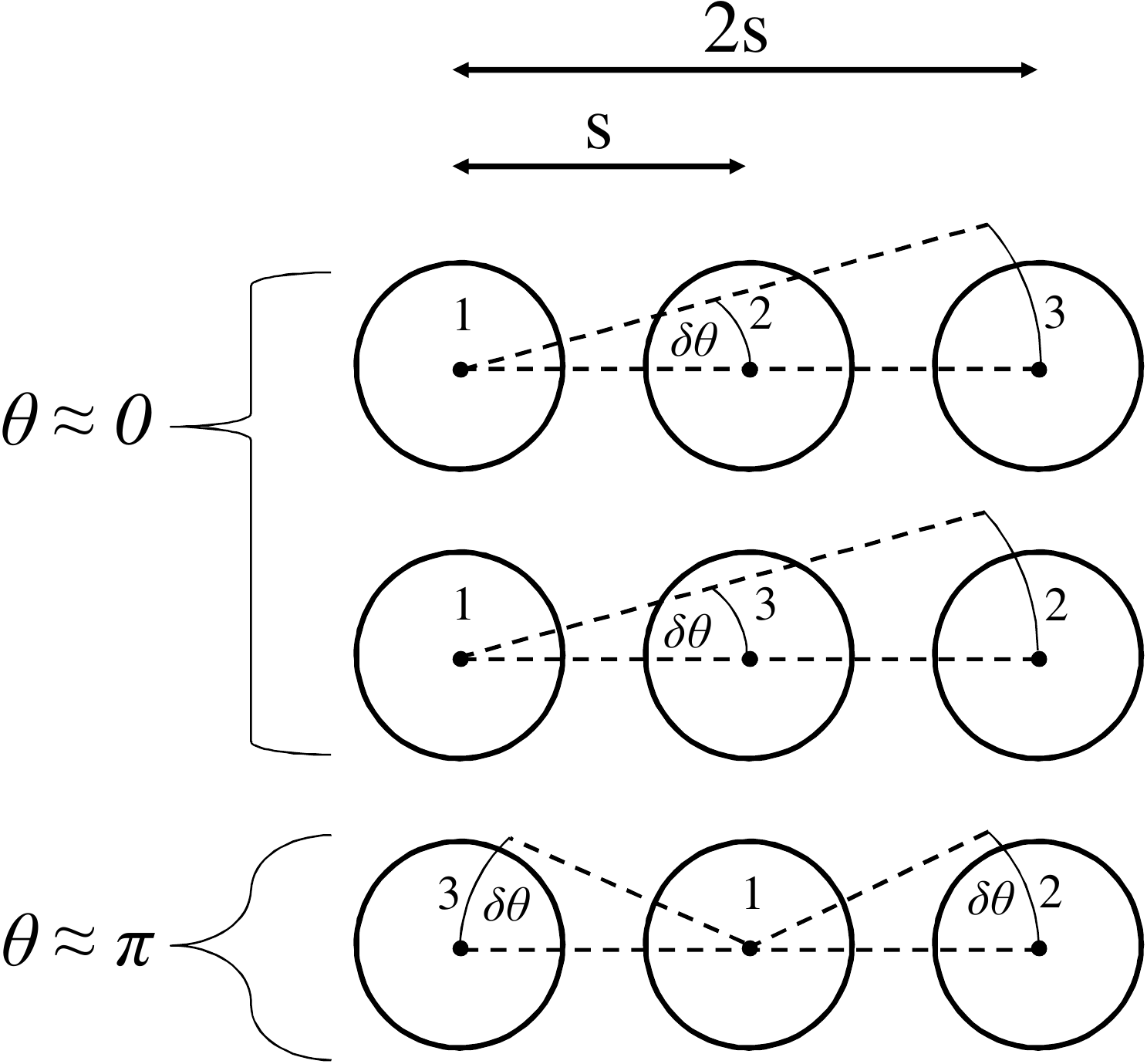}}
\caption{Configurations where $\theta \approx 0$ (top and middle) and $\theta \approx \pi$ (bottom).}
\label{fig:angle4}
\end{figure}

Here we give an explanation as to why $n(0) = 4n(\pi)$.  Shown in Fig.~\ref{fig:angle4} are configurations where $\theta \approx 0$ (top and middle) and $\theta \approx \pi$ (bottom).  At first glance, it would appear that $n(0) = 2n(\pi)$.  However, one must consider how the number of states changes in the vicinity of 0 and $\pi$.  Thus we are interested in 
\begin{eqnarray}\label{dtlimits}
\lim \limits_{\delta \theta \rightarrow 0}\frac{n(\delta \theta)}{n(\pi - \delta \theta)}
\end{eqnarray}

In the top of Fig.~\ref{fig:angle4}, disks 1 and 2 are separated by an average distance $s$ and disks 1 and 3 are separated by $2s$.  The number of states where $\theta \approx 0$ is proportional to the product of the arc lengths subtended by $\delta \theta$ for disks 2 and 3.  Therefore, for the top configuration, $n_{\rm top} \propto (s\delta \theta)(2s \delta\theta) = 2s^2 \delta \theta^2$.  In the same way for the middle configuration, $n_{\rm mid} = n_{\rm top}$.  Therefore, $n(\delta \theta) = 2n_{\rm top} \propto 4s^2 \delta \theta^2$.  For the bottom configuration where $\theta \approx \pi$, the number of states is again proportional to the product of the arc lengths, but in this case  the distance between disks 1 and 3 and disks 1 and 2 is $s$.  Therefore, $n(\pi - \delta \theta) = n_{\rm bot} \propto s^2 \delta \theta^2$.  Thus, the ratio in Eqn.~(\ref{dtlimits}) is equal to 4.

%% file: threecirc5.bbl
\begin{thebibliography}{10}%
\makeatletter
\providecommand \@ifxundefined [1]{%
 \ifx #1\undefined \expandafter \@firstoftwo
 \else \expandafter \@secondoftwo
\fi
}%
\providecommand \@ifnum [1]{%
 \ifnum #1\expandafter \@firstoftwo
 \else \expandafter \@secondoftwo
\fi
}%
\providecommand \enquote [1]{``#1''}%
\providecommand \bibnamefont  [1]{#1}%
\providecommand \bibfnamefont [1]{#1}%
\providecommand \citenamefont [1]{#1}%
\providecommand\href[0]{\@sanitize\@href}%
\providecommand\@href[1]{\endgroup\@@startlink{#1}\endgroup\@@href}%
\providecommand\@@href[1]{#1\@@endlink}%
\providecommand \@sanitize [0]{\begingroup\catcode`\&12\catcode`\#12\relax}%
\@ifxundefined \pdfoutput {\@firstoftwo}{%
 \@ifnum{\z@=\pdfoutput}{\@firstoftwo}{\@secondoftwo}%
}{%
 \providecommand\@@startlink[1]{\leavevmode\special{html:<a href="#1">}}%
 \providecommand\@@endlink[0]{\special{html:</a>}}%
}{%
 \providecommand\@@startlink[1]{%
  \leavevmode
  \pdfstartlink
   attr{/Border[0 0 1 ]/H/I/C[0 1 1]}%
   user{/Subtype/Link/A<</Type/Action/S/URI/URI(#1)>>}%
  \relax
 }%
 \providecommand\@@endlink[0]{\pdfendlink}%
}%
\providecommand \url  [0]{\begingroup\@sanitize \@url }%
\providecommand \@url [1]{\endgroup\@href {#1}{\urlprefix}}%
\providecommand \urlprefix [0]{URL }%
\providecommand \Eprint[0]{\href }%
\@ifxundefined \urlstyle {%
  \providecommand \doi [1]{doi:\discretionary{}{}{}#1}%
}{%
  \providecommand \doi [0]{doi:\discretionary{}{}{}\begingroup
  \urlstyle{rm}\Url }%
}%
\providecommand \doibase [0]{http://dx.doi.org/}%
\providecommand \Doi[1]{\href{\doibase#1}}%
\providecommand \bibAnnote [3]{%
  \BibitemShut{#1}%
  \begin{quotation}\noindent
    \textsc{Key:}\ #2\\\textsc{Annotation:}\ #3%
  \end{quotation}%
}%
\providecommand \bibAnnoteFile [2]{%
  \IfFileExists{#2}{\bibAnnote {#1} {#2} {\input{#2}}}{}%
}%
\providecommand \typeout [0]{\immediate \write \m@ne }%
\providecommand \selectlanguage [0]{\@gobble}%
\providecommand \bibinfo [0]{\@secondoftwo}%
\providecommand \bibfield [0]{\@secondoftwo}%
\providecommand \translation [1]{[#1]}%
\providecommand \BibitemOpen[0]{}%
\providecommand \bibitemStop [0]{}%
\providecommand \bibitemNoStop [0]{.\EOS\space}%
\providecommand \EOS [0]{\spacefactor3000\relax}%
\providecommand \BibitemShut [1]{\csname bibitem#1\endcsname}%
\bibitem{goldstein1969}%
  \BibitemOpen
  \bibfield{author}{%
  \bibinfo {author} {\bibfnamefont{M.}~\bibnamefont{Goldstein}},\ }%
  \bibfield{journal}{%
  \Doi{10.1063/1.1672587}{\bibinfo {journal} {J. Chem. Phys.}}\ }%
  \textbf{\bibinfo {volume} {51}},\ \bibinfo {pages} {3728} (\bibinfo {year}
  {1969})%
  \bibAnnoteFile{NoStop}{goldstein1969}%
\bibitem{stillinger1995}%
  \BibitemOpen
  \bibfield{author}{%
  \bibinfo {author} {\bibfnamefont{F.~H.}\ \bibnamefont{Stillinger}},\ }%
  \bibfield{journal}{%
  \Doi{10.1126/science.267.5206.1935}{\bibinfo {journal} {Science}}\ }%
  \textbf{\bibinfo {volume} {267}},\ \bibinfo {pages} {1935} (\bibinfo {year}
  {1995})%
  \bibAnnoteFile{NoStop}{stillinger1995}%
\bibitem{sciortino05}%
  \BibitemOpen
  \bibfield{author}{%
  \bibinfo {author} {\bibfnamefont{F.}~\bibnamefont{Sciortino}},\ }%
  \bibfield{journal}{%
  \Doi{10.1088/1742-5468/2005/05/P05015}{\bibinfo {journal} {J. Stat. Mech.}}\
  }%
  \textbf{\bibinfo {volume} {2005}},\ \bibinfo {pages} {P05015} (\bibinfo
  {year} {2005})%
  \bibAnnoteFile{NoStop}{sciortino05}%
\bibitem{sastry98}%
  \BibitemOpen
  \bibfield{author}{%
  \bibinfo {author} {\bibfnamefont{S.}~\bibnamefont{Sastry}}, \bibinfo {author}
  {\bibfnamefont{P.~G.}\ \bibnamefont{Debenedetti}},\ and\ \bibinfo {author}
  {\bibfnamefont{F.~H.}\ \bibnamefont{Stillinger}},\ }%
  \bibfield{journal}{%
  \Doi{10.1038/31189}{\bibinfo {journal} {Nature}}\ }%
  \textbf{\bibinfo {volume} {393}},\ \bibinfo {pages} {554} (\bibinfo {year}
  {1998})%
  \bibAnnoteFile{NoStop}{sastry98}%
\bibitem{dasgupta99}%
  \BibitemOpen
  \bibfield{author}{%
  \bibinfo {author} {\bibfnamefont{C.}~\bibnamefont{Dasgupta}}\ and\ \bibinfo
  {author} {\bibfnamefont{O.~T.}\ \bibnamefont{Valls}},\ }%
  \bibfield{journal}{%
  \Doi{10.1103/PhysRevE.59.3123}{\bibinfo {journal} {Phys. Rev. E}}\ }%
  \textbf{\bibinfo {volume} {59}},\ \bibinfo {pages} {3123} (\bibinfo {year}
  {1999})%
  \bibAnnoteFile{NoStop}{dasgupta99}%
\bibitem{dasgupta00}%
  \BibitemOpen
  \bibfield{author}{%
  \bibinfo {author} {\bibfnamefont{C.}~\bibnamefont{Dasgupta}}\ and\ \bibinfo
  {author} {\bibfnamefont{O.~T.}\ \bibnamefont{Valls}},\ }%
  \bibfield{journal}{%
  \Doi{10.1088/0953-8984/12/29/327}{\bibinfo {journal} {J. Phys.: Cond.
  Matt.}}\ }%
  \textbf{\bibinfo {volume} {12}},\ \bibinfo {pages} {6553} (\bibinfo {year}
  {2000})%
  \bibAnnoteFile{NoStop}{dasgupta00}%
\bibitem{schroder00}%
  \BibitemOpen
  \bibfield{author}{%
  \bibinfo {author} {\bibfnamefont{T.~B.}\ \bibnamefont{Schr{\o}~der}},
  \bibinfo {author} {\bibfnamefont{S.}~\bibnamefont{Sastry}}, \bibinfo {author}
  {\bibfnamefont{J.~C.}\ \bibnamefont{Dyre}},\ and\ \bibinfo {author}
  {\bibfnamefont{S.~C.}\ \bibnamefont{Glotzer}},\ }%
  \bibfield{journal}{%
  \Doi{10.1063/1.481621}{\bibinfo {journal} {J. Chem. Phys.}}\ }%
  \textbf{\bibinfo {volume} {112}},\ \bibinfo {pages} {9834} (\bibinfo {year}
  {2000})%
  \bibAnnoteFile{NoStop}{schroder00}%
\bibitem{sastry01}%
  \BibitemOpen
  \bibfield{author}{%
  \bibinfo {author} {\bibfnamefont{S.}~\bibnamefont{Sastry}},\ }%
  \bibfield{journal}{%
  \Doi{10.1038/35051524}{\bibinfo {journal} {Nature}}\ }%
  \textbf{\bibinfo {volume} {409}},\ \bibinfo {pages} {164} (\bibinfo {year}
  {2001})%
  \bibAnnoteFile{NoStop}{sastry01}%
\bibitem{grigera02}%
  \BibitemOpen
  \bibfield{author}{%
  \bibinfo {author} {\bibfnamefont{T.~S.}\ \bibnamefont{Grigera}}, \bibinfo
  {author} {\bibfnamefont{A.}~\bibnamefont{Cavagna}}, \bibinfo {author}
  {\bibfnamefont{I.}~\bibnamefont{Giardina}},\ and\ \bibinfo {author}
  {\bibfnamefont{G.}~\bibnamefont{Parisi}},\ }%
  \bibfield{journal}{%
  \Doi{10.1103/PhysRevLett.88.055502}{\bibinfo {journal} {Phys. Rev. Lett.}}\
  }%
  \textbf{\bibinfo {volume} {88}},\ \bibinfo {pages} {055502} (\bibinfo {year}
  {2002})%
  \bibAnnoteFile{NoStop}{grigera02}%
\bibitem{kim2003}%
  \BibitemOpen
  \bibfield{author}{%
  \bibinfo {author} {\bibfnamefont{K.}~\bibnamefont{Kim}}\ and\ \bibinfo
  {author} {\bibfnamefont{T.}~\bibnamefont{Munakata}},\ }%
  \bibfield{journal}{%
  \Doi{10.1103/PhysRevE.68.021502}{\bibinfo {journal} {Phys. Rev. E}}\ }%
  \textbf{\bibinfo {volume} {68}},\ \bibinfo {pages} {021502} (\bibinfo {year}
  {2003})%
  \bibAnnoteFile{NoStop}{kim2003}%
\bibitem{vogel04}%
  \BibitemOpen
  \bibfield{author}{%
  \bibinfo {author} {\bibfnamefont{M.}~\bibnamefont{Vogel}}, \bibinfo {author}
  {\bibfnamefont{B.}~\bibnamefont{Doliwa}}, \bibinfo {author}
  {\bibfnamefont{A.}~\bibnamefont{Heuer}},\ and\ \bibinfo {author}
  {\bibfnamefont{S.~C.}\ \bibnamefont{Glotzer}},\ }%
  \bibfield{journal}{%
  \Doi{10.1063/1.1644538}{\bibinfo {journal} {J. Chem. Phys.}}\ }%
  \textbf{\bibinfo {volume} {120}},\ \bibinfo {pages} {4404} (\bibinfo {year}
  {2004})%
  \bibAnnoteFile{NoStop}{vogel04}%
\bibitem{schweizerjcp}%
  \BibitemOpen
  \bibfield{author}{%
  \bibinfo {author} {\bibfnamefont{K.~S.}\ \bibnamefont{Schweizer}}\ and\
  \bibinfo {author} {\bibfnamefont{E.~J.}\ \bibnamefont{Saltzman}},\ }%
  \bibfield{journal}{%
  \Doi{10.1063/1.1578632}{\bibinfo {journal} {J. Chem. Phys.}}\ }%
  \textbf{\bibinfo {volume} {119}},\ \bibinfo {pages} {1181} (\bibinfo {year}
  {2003})%
  \bibAnnoteFile{NoStop}{schweizerjcp}%
\bibitem{bonn07}%
  \BibitemOpen
  \bibfield{author}{%
  \bibinfo {author} {\bibfnamefont{S.~J.}\ \bibnamefont{Farouji}}, \bibinfo
  {author} {\bibfnamefont{G.~H.}\ \bibnamefont{Wegdam}},\ and\ \bibinfo
  {author} {\bibfnamefont{D.}~\bibnamefont{Bonn}},\ }%
  \bibfield{journal}{%
  \Doi{10.1103/PhysRevLett.99.065701}{\bibinfo {journal} {Phys. Rev. Lett.}}\
  }%
  \textbf{\bibinfo {volume} {99}},\ \bibinfo {pages} {065701} (\bibinfo {year}
  {2007})%
  \bibAnnoteFile{NoStop}{bonn07}%
\bibitem{kauzmann48}%
  \BibitemOpen
  \bibfield{author}{%
  \bibinfo {author} {\bibfnamefont{W.}~\bibnamefont{Kauzmann}},\ }%
  \bibfield{journal}{%
  \Doi{10.1021/cr60135a002}{\bibinfo {journal} {Chem. Rev.}}\ }%
  \textbf{\bibinfo {volume} {43}},\ \bibinfo {pages} {219} (\bibinfo {year}
  {1948})%
  \bibAnnoteFile{NoStop}{kauzmann48}%
\bibitem{debenedetti2001nat}%
  \BibitemOpen
  \bibfield{author}{%
  \bibinfo {author} {\bibfnamefont{P.~G.}\ \bibnamefont{Debenedetti}}\ and\
  \bibinfo {author} {\bibfnamefont{F.~H.}\ \bibnamefont{Stillinger}},\ }%
  \bibfield{journal}{%
  \Doi{10.1038/35065704}{\bibinfo {journal} {Nature}}\ }%
  \textbf{\bibinfo {volume} {410}},\ \bibinfo {pages} {259} (\bibinfo {year}
  {2001})%
  \bibAnnoteFile{NoStop}{debenedetti2001nat}%
\bibitem{walesbook}%
  \BibitemOpen
  \bibfield{author}{%
  \bibinfo {author} {\bibfnamefont{D.~J.}\ \bibnamefont{Wales}},\ }%
  \emph{\bibinfo {title} {Energy Landscapes: Applications to Clusters,
  Biomolecules and Glasses (Cambridge Molecular Science)}},\ \bibinfo {edition}
  {1st}\ ed.\ (\bibinfo {publisher} {Cambridge University Press},\ \bibinfo
  {year} {2004})%
  \bibAnnoteFile{NoStop}{walesbook}%
\bibitem{frenkel92}%
  \BibitemOpen
  \bibfield{author}{%
  \bibinfo {author} {\bibfnamefont{J.~S.}\ \bibnamefont{van Duijneveldt}}\ and\
  \bibinfo {author} {\bibfnamefont{D.}~\bibnamefont{Frenkel}},\ }%
  \bibfield{journal}{%
  \Doi{10.1063/1.462802}{\bibinfo {journal} {J. Chem. Phys.}}\ }%
  \textbf{\bibinfo {volume} {96}},\ \bibinfo {pages} {4655} (\bibinfo {year}
  {1992})%
  \bibAnnoteFile{NoStop}{frenkel92}%
\bibitem{frenkel96}%
  \BibitemOpen
  \bibfield{author}{%
  \bibinfo {author} {\bibfnamefont{P.~R.}\ \bibnamefont{ten Wolde}}, \bibinfo
  {author} {\bibfnamefont{M.~J.}\ \bibnamefont{Ruiz~Montero}},\ and\ \bibinfo
  {author} {\bibfnamefont{D.}~\bibnamefont{Frenkel}},\ }%
  \bibfield{journal}{%
  \Doi{10.1063/1.471721}{\bibinfo {journal} {J. Chem. Phys.}}\ }%
  \textbf{\bibinfo {volume} {104}},\ \bibinfo {pages} {9932} (\bibinfo {year}
  {1996})%
  \bibAnnoteFile{NoStop}{frenkel96}%
\bibitem{frenkel97}%
  \BibitemOpen
  \bibfield{author}{%
  \bibinfo {author} {\bibfnamefont{P.~R.~T.}\ \bibnamefont{Wolde}}\ and\
  \bibinfo {author} {\bibfnamefont{D.}~\bibnamefont{Frenkel}},\ }%
  \bibfield{journal}{%
  \Doi{10.1126/science.277.5334.1975}{\bibinfo {journal} {Science}}\ }%
  \textbf{\bibinfo {volume} {277}},\ \bibinfo {pages} {1975} (\bibinfo {year}
  {1997})%
  \bibAnnoteFile{NoStop}{frenkel97}%
\bibitem{heuer97}%
  \BibitemOpen
  \bibfield{author}{%
  \bibinfo {author} {\bibfnamefont{A.}~\bibnamefont{Heuer}},\ }%
  \bibfield{journal}{%
  \Doi{10.1103/PhysRevLett.78.4051}{\bibinfo {journal} {Phys. Rev. Lett.}}\ }%
  \textbf{\bibinfo {volume} {78}},\ \bibinfo {pages} {4051} (\bibinfo {year}
  {1997})%
  \bibAnnoteFile{NoStop}{heuer97}%
\bibitem{krivov04}%
  \BibitemOpen
  \bibfield{author}{%
  \bibinfo {author} {\bibfnamefont{S.~V.}\ \bibnamefont{Krivov}}\ and\ \bibinfo
  {author} {\bibfnamefont{M.}~\bibnamefont{Karplus}},\ }%
  \bibfield{journal}{%
  \Doi{10.1073/pnas.0406234101}{\bibinfo {journal} {Proc. Nat. Acad. Sci.}}\ }%
  \textbf{\bibinfo {volume} {101}},\ \bibinfo {pages} {14766} (\bibinfo {year}
  {2004})%
  \bibAnnoteFile{NoStop}{krivov04}%
\bibitem{altis08}%
  \BibitemOpen
  \bibfield{author}{%
  \bibinfo {author} {\bibfnamefont{A.}~\bibnamefont{Altis}}, \bibinfo {author}
  {\bibfnamefont{M.}~\bibnamefont{Otten}}, \bibinfo {author}
  {\bibfnamefont{P.~H.}\ \bibnamefont{Nguyen}}, \bibinfo {author}
  {\bibfnamefont{R.}~\bibnamefont{Hegger}},\ and\ \bibinfo {author}
  {\bibfnamefont{G.}~\bibnamefont{Stock}},\ }%
  \bibfield{journal}{%
  \Doi{10.1063/1.2945165}{\bibinfo {journal} {J. Chem. Phys.}}\ }%
  \textbf{\bibinfo {volume} {128}},\ \bibinfo {pages} {245102} (\bibinfo {year}
  {2008})%
  \bibAnnoteFile{NoStop}{altis08}%
\bibitem{liwo09}%
  \BibitemOpen
  \bibfield{author}{%
  \bibinfo {author} {\bibfnamefont{G.~G.}\ \bibnamefont{Maisuradze}}, \bibinfo
  {author} {\bibfnamefont{A.}~\bibnamefont{Liwo}},\ and\ \bibinfo {author}
  {\bibfnamefont{H.~A.}\ \bibnamefont{Scheraga}},\ }%
  \bibfield{journal}{%
  \Doi{10.1103/PhysRevLett.102.238102}{\bibinfo {journal} {Phys. Rev. Lett.}}\
  }%
  \textbf{\bibinfo {volume} {102}},\ \bibinfo {pages} {238102} (\bibinfo {year}
  {2009})%
  \bibAnnoteFile{NoStop}{liwo09}%
\bibitem{bevan11}%
  \BibitemOpen
  \bibfield{author}{%
  \bibinfo {author} {\bibfnamefont{M.~A.}\ \bibnamefont{Bevan}},\ }%
  \bibfield{journal}{%
  \Doi{10.1039/c0sm01526a}{\bibinfo {journal} {Soft Matter}}\ }%
  \textbf{\bibinfo {volume} {7}},\ \bibinfo {pages} {3280} (\bibinfo {year}
  {2011})%
  \bibAnnoteFile{NoStop}{bevan11}%
\bibitem{lowenhardspheres}%
  \BibitemOpen
  \bibfield{author}{%
  \bibinfo {author} {\bibfnamefont{H.}~\bibnamefont{L\"{o}wen}},\ }%
  \enquote{\bibinfo {title} {Fun with hard spheres},}\ in\ \emph{\bibinfo
  {booktitle} {Statistical Physics and Spatial Statistics (Lecture Notes in
  Physics)}},\ Vol.\ \bibinfo {volume} {554},\ \bibinfo {editor} {edited by\
  \bibinfo {editor} {\bibfnamefont{K.~R.}\ \bibnamefont{Mecke}}\ and\ \bibinfo
  {editor} {\bibfnamefont{D.}~\bibnamefont{Stoyan}}}\ (\bibinfo {publisher}
  {Berlin: Springer},\ \bibinfo {year} {2000})\ pp.\ \bibinfo {pages}
  {295--331}%
  \bibAnnoteFile{NoStop}{lowenhardspheres}%
\bibitem{donev07}%
  \BibitemOpen
  \bibfield{author}{%
  \bibinfo {author} {\bibfnamefont{A.}~\bibnamefont{Donev}}, \bibinfo {author}
  {\bibfnamefont{F.~H.}\ \bibnamefont{Stillinger}},\ and\ \bibinfo {author}
  {\bibfnamefont{S.}~\bibnamefont{Torquato}},\ }%
  \bibfield{journal}{%
  \Doi{10.1063/1.2775928}{\bibinfo {journal} {J. Chem. Phys.}}\ }%
  \textbf{\bibinfo {volume} {127}},\ \bibinfo {pages} {124509} (\bibinfo {year}
  {2007})%
  \bibAnnoteFile{NoStop}{donev07}%
\bibitem{speedytwodiscs}%
  \BibitemOpen
  \bibfield{author}{%
  \bibinfo {author} {\bibfnamefont{R.~J.}\ \bibnamefont{Speedy}},\ }%
  \bibfield{journal}{%
  \Doi{10.1016/0378-4371(94)90082-5}{\bibinfo {journal} {Physica A}}\ }%
  \textbf{\bibinfo {volume} {210}},\ \bibinfo {pages} {341} (\bibinfo {year}
  {1994})%
  \bibAnnoteFile{NoStop}{speedytwodiscs}%
\bibitem{speedy1999discs}%
  \BibitemOpen
  \bibfield{author}{%
  \bibinfo {author} {\bibfnamefont{R.~K.}\ \bibnamefont{Bowles}}\ and\ \bibinfo
  {author} {\bibfnamefont{R.~J.}\ \bibnamefont{Speedy}},\ }%
  \bibfield{journal}{%
  \Doi{10.1016/S0378-4371(98)00404-X}{\bibinfo {journal} {Physica A}}\ }%
  \textbf{\bibinfo {volume} {262}},\ \bibinfo {pages} {76} (\bibinfo {year}
  {1999})%
  \bibAnnoteFile{NoStop}{speedy1999discs}%
\bibitem{schweizerjpcb}%
  \BibitemOpen
  \bibfield{author}{%
  \bibinfo {author} {\bibfnamefont{K.~S.}\ \bibnamefont{Schweizer}}\ and\
  \bibinfo {author} {\bibfnamefont{E.~J.}\ \bibnamefont{Saltzman}},\ }%
  \bibfield{journal}{%
  \Doi{10.1021/jp047763j}{\bibinfo {journal} {J. Phys. Chem. B}}\ }%
  \textbf{\bibinfo {volume} {108}},\ \bibinfo {pages} {19729} (\bibinfo {year}
  {2004})%
  \bibAnnoteFile{NoStop}{schweizerjpcb}%
\bibitem{berne66}%
  \BibitemOpen
  \bibfield{author}{%
  \bibinfo {author} {\bibfnamefont{B.~J.}\ \bibnamefont{Berne}}, \bibinfo
  {author} {\bibfnamefont{J.~P.}\ \bibnamefont{Boon}},\ and\ \bibinfo {author}
  {\bibfnamefont{S.~A.}\ \bibnamefont{Rice}},\ }%
  \bibfield{journal}{%
  \Doi{10.1063/1.1727719}{\bibinfo {journal} {J. Chem. Phys.}}\ }%
  \textbf{\bibinfo {volume} {45}},\ \bibinfo {pages} {1086} (\bibinfo {year}
  {1966})%
  \bibAnnoteFile{NoStop}{berne66}%
\bibitem{sjogren80}%
  \BibitemOpen
  \bibfield{author}{%
  \bibinfo {author} {\bibfnamefont{L.}~\bibnamefont{Sj\"{o}gren}},\ }%
  \bibfield{journal}{%
  \Doi{10.1103/PhysRevA.22.2883}{\bibinfo {journal} {Phys. Rev. A}}\ }%
  \textbf{\bibinfo {volume} {22}},\ \bibinfo {pages} {2883} (\bibinfo {year}
  {1980})%
  \bibAnnoteFile{NoStop}{sjogren80}%
\bibitem{wahnstrom82}%
  \BibitemOpen
  \bibfield{author}{%
  \bibinfo {author} {\bibfnamefont{G.}~\bibnamefont{Wahnstrom}}\ and\ \bibinfo
  {author} {\bibfnamefont{L.}~\bibnamefont{Sj{\"o}gren}},\ }%
  \bibfield{journal}{%
  \Doi{10.1088/0022-3719/15/3/007}{\bibinfo {journal} {J. Phys. C: Solid State
  Physics}}\ }%
  \textbf{\bibinfo {volume} {15}},\ \bibinfo {pages} {401} (\bibinfo {year}
  {1982})%
  \bibAnnoteFile{NoStop}{wahnstrom82}%
\bibitem{gotze1992rpp}%
  \BibitemOpen
  \bibfield{author}{%
  \bibinfo {author} {\bibfnamefont{W.}~\bibnamefont{Gotze}}\ and\ \bibinfo
  {author} {\bibfnamefont{L.}~\bibnamefont{Sj{\"o}gren}},\ }%
  \bibfield{journal}{%
  \Doi{10.1088/0034-4885/55/3/001}{\bibinfo {journal} {Rep. Prog. Phys.}}\ }%
  \textbf{\bibinfo {volume} {55}},\ \bibinfo {pages} {241} (\bibinfo {year}
  {1992})%
  \bibAnnoteFile{NoStop}{gotze1992rpp}%
\bibitem{doliwa1998prl}%
  \BibitemOpen
  \bibfield{author}{%
  \bibinfo {author} {\bibfnamefont{B.}~\bibnamefont{Doliwa}}\ and\ \bibinfo
  {author} {\bibfnamefont{A.}~\bibnamefont{Heuer}},\ }%
  \bibfield{journal}{%
  \Doi{10.1103/PhysRevLett.80.4915}{\bibinfo {journal} {Phys. Rev. Lett.}}\ }%
  \textbf{\bibinfo {volume} {80}},\ \bibinfo {pages} {4915} (\bibinfo {year}
  {1998})%
  \bibAnnoteFile{NoStop}{doliwa1998prl}%
\bibitem{weeks02sub}%
  \BibitemOpen
  \bibfield{author}{%
  \bibinfo {author} {\bibfnamefont{E.~R.}\ \bibnamefont{Weeks}}\ and\ \bibinfo
  {author} {\bibfnamefont{D.~A.}\ \bibnamefont{Weitz}},\ }%
  \bibfield{journal}{%
  \Doi{10.1016/S0301-0104(02)00667-5}{\bibinfo {journal} {Chem. Phys.}}\ }%
  \textbf{\bibinfo {volume} {284}},\ \bibinfo {pages} {361} (\bibinfo {year}
  {2002})%
  \bibAnnoteFile{NoStop}{weeks02sub}%
\bibitem{weeks02}%
  \BibitemOpen
  \bibfield{author}{%
  \bibinfo {author} {\bibfnamefont{E.~R.}\ \bibnamefont{Weeks}}\ and\ \bibinfo
  {author} {\bibfnamefont{D.~A.}\ \bibnamefont{Weitz}},\ }%
  \bibfield{journal}{%
  \Doi{10.1103/PhysRevLett.89.095704}{\bibinfo {journal} {Phys. Rev. Lett.}}\
  }%
  \textbf{\bibinfo {volume} {89}},\ \bibinfo {pages} {095704} (\bibinfo {year}
  {2002})%
  \bibAnnoteFile{NoStop}{weeks02}%
\bibitem{kvlee2004}%
  \BibitemOpen
  \bibfield{author}{%
  \bibinfo {author} {\bibfnamefont{K.}~\bibnamefont{Vollmayr-Lee}},\ }%
  \bibfield{journal}{%
  \Doi{10.1063/1.1778155}{\bibinfo {journal} {J. Chem. Phys.}}\ }%
  \textbf{\bibinfo {volume} {121}},\ \bibinfo {pages} {4781} (\bibinfo {year}
  {2004})%
  \bibAnnoteFile{NoStop}{kvlee2004}%
\bibitem{sirono2011}%
  \BibitemOpen
  \bibfield{author}{%
  \bibinfo {author} {\bibfnamefont{S.~I.}\ \bibnamefont{Sirono}},\ }%
  \bibfield{journal}{%
  \Doi{10.1103/PhysRevB.84.104201}{\bibinfo {journal} {Phys. Rev. B}}\ }%
  \textbf{\bibinfo {volume} {84}},\ \bibinfo {pages} {104201} (\bibinfo {year}
  {2011})%
  \bibAnnoteFile{NoStop}{sirono2011}%
\bibitem{wales2008}%
  \BibitemOpen
  \bibfield{author}{%
  \bibinfo {author} {\bibfnamefont{V.~K.}\ \bibnamefont{de~Souza}}\ and\
  \bibinfo {author} {\bibfnamefont{D.~J.}\ \bibnamefont{Wales}},\ }%
  \bibfield{journal}{%
  \Doi{10.1063/1.2992128}{\bibinfo {journal} {J. Chem. Phys.}}\ }%
  \textbf{\bibinfo {volume} {129}},\ \bibinfo {pages} {164507} (\bibinfo {year}
  {2008})%
  \bibAnnoteFile{NoStop}{wales2008}%
\bibitem{ohern2008}%
  \BibitemOpen
  \bibfield{author}{%
  \bibinfo {author} {\bibfnamefont{P.}~\bibnamefont{Pal}}, \bibinfo {author}
  {\bibfnamefont{C.~S.}\ \bibnamefont{O'Hern}}, \bibinfo {author}
  {\bibfnamefont{J.}~\bibnamefont{Blawzdziewicz}}, \bibinfo {author}
  {\bibfnamefont{E.~R.}\ \bibnamefont{Dufresne}},\ and\ \bibinfo {author}
  {\bibfnamefont{R.}~\bibnamefont{Stinchcombe}},\ }%
  \bibfield{journal}{%
  \Doi{10.1103/PhysRevE.78.011111}{\bibinfo {journal} {Phys. Rev. E}}\ }%
  \textbf{\bibinfo {volume} {78}},\ \bibinfo {pages} {011111} (\bibinfo {year}
  {2008})%
  \bibAnnoteFile{NoStop}{ohern2008}%
\bibitem{alder63}%
  \BibitemOpen
  \bibfield{author}{%
  \bibinfo {author} {\bibfnamefont{B.~J.}\ \bibnamefont{Alder}}, \bibinfo
  {author} {\bibfnamefont{W.~G.}\ \bibnamefont{Hoover}},\ and\ \bibinfo
  {author} {\bibfnamefont{T.~E.}\ \bibnamefont{Wainwright}},\ }%
  \bibfield{journal}{%
  \Doi{10.1103/PhysRevLett.11.241}{\bibinfo {journal} {Phys. Rev. Lett.}}\ }%
  \textbf{\bibinfo {volume} {11}},\ \bibinfo {pages} {241} (\bibinfo {year}
  {1963})%
  \bibAnnoteFile{NoStop}{alder63}%
\bibitem{adam1965jcp}%
  \BibitemOpen
  \bibfield{author}{%
  \bibinfo {author} {\bibfnamefont{G.}~\bibnamefont{Adam}}\ and\ \bibinfo
  {author} {\bibfnamefont{J.~H.}\ \bibnamefont{Gibbs}},\ }%
  \bibfield{journal}{%
  \Doi{10.1063/1.1696442}{\bibinfo {journal} {J. Chem. Phys.}}\ }%
  \textbf{\bibinfo {volume} {43}},\ \bibinfo {pages} {139} (\bibinfo {year}
  {1965})%
  \bibAnnoteFile{NoStop}{adam1965jcp}%
\bibitem{donati1998prl}%
  \BibitemOpen
  \bibfield{author}{%
  \bibinfo {author} {\bibfnamefont{C.}~\bibnamefont{Donati}}, \bibinfo {author}
  {\bibfnamefont{J.~F.}\ \bibnamefont{Douglas}}, \bibinfo {author}
  {\bibfnamefont{W.}~\bibnamefont{Kob}}, \bibinfo {author}
  {\bibfnamefont{S.~J.}\ \bibnamefont{Plimpton}}, \bibinfo {author}
  {\bibfnamefont{P.~H.}\ \bibnamefont{Poole}},\ and\ \bibinfo {author}
  {\bibfnamefont{S.~C.}\ \bibnamefont{Glotzer}},\ }%
  \bibfield{journal}{%
  \Doi{10.1103/PhysRevLett.80.2338}{\bibinfo {journal} {Phys. Rev. Lett.}}\ }%
  \textbf{\bibinfo {volume} {80}},\ \bibinfo {pages} {2338} (\bibinfo {year}
  {1998})%
  \bibAnnoteFile{NoStop}{donati1998prl}%
\bibitem{marcus1999pre}%
  \BibitemOpen
  \bibfield{author}{%
  \bibinfo {author} {\bibfnamefont{A.~H.}\ \bibnamefont{Marcus}}, \bibinfo
  {author} {\bibfnamefont{J.}~\bibnamefont{Schofield}},\ and\ \bibinfo {author}
  {\bibfnamefont{S.~A.}\ \bibnamefont{Rice}},\ }%
  \bibfield{journal}{%
  \Doi{10.1103/PhysRevE.60.5725}{\bibinfo {journal} {Phys. Rev. E}}\ }%
  \textbf{\bibinfo {volume} {60}},\ \bibinfo {pages} {5725} (\bibinfo {year}
  {1999})%
  \bibAnnoteFile{NoStop}{marcus1999pre}%
\bibitem{doliwa2000pre}%
  \BibitemOpen
  \bibfield{author}{%
  \bibinfo {author} {\bibfnamefont{B.}~\bibnamefont{Doliwa}}\ and\ \bibinfo
  {author} {\bibfnamefont{A.}~\bibnamefont{Heuer}},\ }%
  \bibfield{journal}{%
  \Doi{10.1103/PhysRevE.61.6898}{\bibinfo {journal} {Phys. Rev. E}}\ }%
  \textbf{\bibinfo {volume} {61}},\ \bibinfo {pages} {6898} (\bibinfo {year}
  {2000})%
  \bibAnnoteFile{NoStop}{doliwa2000pre}%
\bibitem{scheidler02}%
  \BibitemOpen
  \bibfield{author}{%
  \bibinfo {author} {\bibfnamefont{P.}~\bibnamefont{Scheidler}}, \bibinfo
  {author} {\bibfnamefont{W.}~\bibnamefont{Kob}},\ and\ \bibinfo {author}
  {\bibfnamefont{K.}~\bibnamefont{Binder}},\ }%
  \bibfield{journal}{%
  \Doi{10.1209/epl/i2002-00182-9}{\bibinfo {journal} {Europhys. Lett.}}\ }%
  \textbf{\bibinfo {volume} {59}},\ \bibinfo {pages} {701} (\bibinfo {year}
  {2002})%
  \bibAnnoteFile{NoStop}{scheidler02}%
\bibitem{ediger2000arpc}%
  \BibitemOpen
  \bibfield{author}{%
  \bibinfo {author} {\bibfnamefont{M.~D.}\ \bibnamefont{Ediger}},\ }%
  \bibfield{journal}{%
  \Doi{10.1146/annurev.physchem.51.1.99}{\bibinfo {journal} {Ann. Rev. Phys.
  Chem.}}\ }%
  \textbf{\bibinfo {volume} {51}},\ \bibinfo {pages} {99} (\bibinfo {year}
  {2000})%
  \bibAnnoteFile{NoStop}{ediger2000arpc}%
\bibitem{richert02}%
  \BibitemOpen
  \bibfield{author}{%
  \bibinfo {author} {\bibfnamefont{R.}~\bibnamefont{Richert}},\ }%
  \bibfield{journal}{%
  \Doi{10.1088/0953-8984/14/23/201}{\bibinfo {journal} {J. Phys.: Cond.
  Matt.}}\ }%
  \textbf{\bibinfo {volume} {14}},\ \bibinfo {pages} {R703} (\bibinfo {year}
  {2002})%
  \bibAnnoteFile{NoStop}{richert02}%
\bibitem{sillescu1999}%
  \BibitemOpen
  \bibfield{author}{%
  \bibinfo {author} {\bibfnamefont{H.}~\bibnamefont{Sillescu}},\ }%
  \bibfield{journal}{%
  \Doi{10.1016/S0022-3093(98)00831-X}{\bibinfo {journal} {J. Non-Cryst.
  Solids}}\ }%
  \textbf{\bibinfo {volume} {243}},\ \bibinfo {pages} {81} (\bibinfo {year}
  {1999})%
  \bibAnnoteFile{NoStop}{sillescu1999}%
\bibitem{odagaki06}%
  \BibitemOpen
  \bibfield{author}{%
  \bibinfo {author} {\bibfnamefont{T.}~\bibnamefont{Odagaki}}, \bibinfo
  {author} {\bibfnamefont{T.}~\bibnamefont{Yoshidome}}, \bibinfo {author}
  {\bibfnamefont{A.}~\bibnamefont{Koyama}},\ and\ \bibinfo {author}
  {\bibfnamefont{A.}~\bibnamefont{Yoshimori}},\ }%
  \bibfield{journal}{%
  \Doi{10.1016/j.jnoncrysol.2006.02.146}{\bibinfo {journal} {J. Non-Cryst.
  Solids}}\ }%
  \textbf{\bibinfo {volume} {352}},\ \bibinfo {pages} {4843} (\bibinfo {year}
  {2006})%
  \bibAnnoteFile{NoStop}{odagaki06}%
\bibitem{yoshidome07}%
  \BibitemOpen
  \bibfield{author}{%
  \bibinfo {author} {\bibfnamefont{T.}~\bibnamefont{Yoshidome}}, \bibinfo
  {author} {\bibfnamefont{A.}~\bibnamefont{Yoshimori}},\ and\ \bibinfo {author}
  {\bibfnamefont{T.}~\bibnamefont{Odagaki}},\ }%
  \bibfield{journal}{%
  \Doi{10.1103/PhysRevE.76.021506}{\bibinfo {journal} {Phys. Rev. E}}\ }%
  \textbf{\bibinfo {volume} {76}},\ \bibinfo {pages} {021506} (\bibinfo {year}
  {2007})%
  \bibAnnoteFile{NoStop}{yoshidome07}%
\bibitem{yoshidome08}%
  \BibitemOpen
  \bibfield{author}{%
  \bibinfo {author} {\bibfnamefont{T.}~\bibnamefont{Yoshidome}}, \bibinfo
  {author} {\bibfnamefont{T.}~\bibnamefont{Odagaki}},\ and\ \bibinfo {author}
  {\bibfnamefont{A.}~\bibnamefont{Yoshimori}},\ }%
  \bibfield{journal}{%
  \Doi{10.1103/PhysRevE.77.061503}{\bibinfo {journal} {Phys. Rev. E}}\ }%
  \textbf{\bibinfo {volume} {77}},\ \bibinfo {pages} {061503} (\bibinfo {year}
  {2008})%
  \bibAnnoteFile{NoStop}{yoshidome08}%
\bibitem{heuer00}%
  \BibitemOpen
  \bibfield{author}{%
  \bibinfo {author} {\bibfnamefont{S.}~\bibnamefont{B\"{u}chner}}\ and\
  \bibinfo {author} {\bibfnamefont{A.}~\bibnamefont{Heuer}},\ }%
  \bibfield{journal}{%
  \Doi{10.1103/PhysRevLett.84.2168}{\bibinfo {journal} {Phys. Rev. Lett.}}\ }%
  \textbf{\bibinfo {volume} {84}},\ \bibinfo {pages} {2168} (\bibinfo {year}
  {2000})%
  \bibAnnoteFile{NoStop}{heuer00}%
\bibitem{segrepusey97}%
  \BibitemOpen
  \bibfield{author}{%
  \bibinfo {author} {\bibfnamefont{P.}~\bibnamefont{Pusey}}, \bibinfo {author}
  {\bibfnamefont{P.}~\bibnamefont{Segr{\`{e}}}}, \bibinfo {author}
  {\bibfnamefont{O.}~\bibnamefont{Behrend}}, \bibinfo {author}
  {\bibfnamefont{S.}~\bibnamefont{Meeker}},\ and\ \bibinfo {author}
  {\bibfnamefont{W.}~\bibnamefont{Poon}},\ }%
  \bibfield{journal}{%
  \Doi{10.1016/S0378-4371(96)00323-8}{\bibinfo {journal} {Physica A}}\ }%
  \textbf{\bibinfo {volume} {235}},\ \bibinfo {pages} {1} (\bibinfo {year}
  {1997})%
  \bibAnnoteFile{NoStop}{segrepusey97}%
\bibitem{pusey2008jpcm}%
  \BibitemOpen
  \bibfield{author}{%
  \bibinfo {author} {\bibfnamefont{P.~N.}\ \bibnamefont{Pusey}},\ }%
  \bibfield{journal}{%
  \Doi{10.1088/0953-8984/20/49/494202}{\bibinfo {journal} {J. Phys.: Cond.
  Matt.}}\ }%
  \textbf{\bibinfo {volume} {20}},\ \bibinfo {pages} {494202} (\bibinfo {year}
  {2008})%
  \bibAnnoteFile{NoStop}{pusey2008jpcm}%
\bibitem{cianci2006ssc}%
  \BibitemOpen
  \bibfield{author}{%
  \bibinfo {author} {\bibfnamefont{G.~C.}\ \bibnamefont{Cianci}}, \bibinfo
  {author} {\bibfnamefont{R.~E.}\ \bibnamefont{Courtland}},\ and\ \bibinfo
  {author} {\bibfnamefont{E.~R.}\ \bibnamefont{Weeks}},\ }%
  \bibfield{journal}{%
  \Doi{10.1016/j.ssc.2006.04.039}{\bibinfo {journal} {Solid State Comm.}}\ }%
  \textbf{\bibinfo {volume} {139}},\ \bibinfo {pages} {599} (\bibinfo {year}
  {2006})%
  \bibAnnoteFile{NoStop}{cianci2006ssc}%
\bibitem{vanBlaaderen1995}%
  \BibitemOpen
  \bibfield{author}{%
  \bibinfo {author} {\bibfnamefont{A.}~\bibnamefont{van Blaaderen}}\ and\
  \bibinfo {author} {\bibfnamefont{P.}~\bibnamefont{Wiltzius}},\ }%
  \bibfield{journal}{%
  \Doi{10.1126/science.270.5239.1177}{\bibinfo {journal} {Science}}\ }%
  \textbf{\bibinfo {volume} {270}},\ \bibinfo {pages} {1177} (\bibinfo {year}
  {1995})%
  \bibAnnoteFile{NoStop}{vanBlaaderen1995}%
\bibitem{frenkelbook}%
  \BibitemOpen
  \bibfield{author}{%
  \bibinfo {author} {\bibfnamefont{D.}~\bibnamefont{Frenkel}}\ and\ \bibinfo
  {author} {\bibfnamefont{B.}~\bibnamefont{Smit}},\ }%
  \Doi{10.1063/1.881812}{\emph{\bibinfo {title} {Understanding Molecular
  Simulation, Second Edition: From Algorithms to Applications (Computational
  Science)}}},\ \bibinfo {edition} {2nd}\ ed.\ (\bibinfo {publisher} {Academic
  Press},\ \bibinfo {year} {2001})%
  \bibAnnoteFile{NoStop}{frenkelbook}%
\bibitem{threecircfootnote1}%
  \BibitemOpen
  \emph{\bibinfo {title} {{\rm While $n(h)$ should have mirror symmetry about
  the line $h = 0$, an artifact of the finite number of Monte Carlo steps is
  that directly computed distributions may show slight unphysical biases toward
  either side of $h = 0$. Though very small, we eliminate these biases by
  averaging one side of a histogram with the mirror image of the other}}}%
  \bibAnnoteFile{NoStop}{threecircfootnote1}%
\bibitem{kramers40}%
  \BibitemOpen
  \bibfield{author}{%
  \bibinfo {author} {\bibfnamefont{H.}~\bibnamefont{Kramers}},\ }%
  \bibfield{journal}{%
  \Doi{10.1016/S0031-8914(40)90098-2}{\bibinfo {journal} {Physica}}\ }%
  \textbf{\bibinfo {volume} {7}},\ \bibinfo {pages} {284} (\bibinfo {year}
  {1940})%
  \bibAnnoteFile{NoStop}{kramers40}%
\bibitem{kramers50}%
  \BibitemOpen
  \bibfield{author}{%
  \bibinfo {author} {\bibfnamefont{P.}~\bibnamefont{H\"{a}nggi}}, \bibinfo
  {author} {\bibfnamefont{P.}~\bibnamefont{Talkner}},\ and\ \bibinfo {author}
  {\bibfnamefont{M.}~\bibnamefont{Borkovec}},\ }%
  \bibfield{journal}{%
  \Doi{10.1103/RevModPhys.62.251}{\bibinfo {journal} {Reviews of Modern
  Physics}}\ }%
  \textbf{\bibinfo {volume} {62}},\ \bibinfo {pages} {251} (\bibinfo {year}
  {1990})%
  \bibAnnoteFile{NoStop}{kramers50}%
\bibitem{gardinerbook}%
  \BibitemOpen
  \bibfield{author}{%
  \bibinfo {author} {\bibfnamefont{C.}~\bibnamefont{Gardiner}},\ }%
  \emph{\bibinfo {title} {Handbook of Stochastic Methods: for Physics,
  Chemistry and the Natural Sciences (Springer Series in Synergetics)}},\
  \bibinfo {edition} {3rd}\ ed.\ (\bibinfo {publisher} {Springer},\ \bibinfo
  {year} {2004})%
  \bibAnnoteFile{NoStop}{gardinerbook}%
\bibitem{threecircfootnote2}%
  \BibitemOpen
  \emph{\bibinfo {title} {\rm{We currently do not have a method to analytically
  calculate $F(\theta)$. Each curve in Fig. \ref{fig:tenvhen}(a) was generated
  by computing $F(\theta)$ from $10^{10}$ configurations of three disks placed
  randomly in the corral and then using Eqn. (\ref{eqn:fcalc}).}}}%
  \bibAnnoteFile{Stop}{threecircfootnote2}%
\bibitem{bryanthardsphere}%
  \BibitemOpen
  \bibfield{author}{%
  \bibinfo {author} {\bibfnamefont{G.}~\bibnamefont{Bryant}}, \bibinfo {author}
  {\bibfnamefont{S.~R.}\ \bibnamefont{Williams}}, \bibinfo {author}
  {\bibfnamefont{L.}~\bibnamefont{Qian}}, \bibinfo {author}
  {\bibfnamefont{I.~K.}\ \bibnamefont{Snook}}, \bibinfo {author}
  {\bibfnamefont{E.}~\bibnamefont{Perez}},\ and\ \bibinfo {author}
  {\bibfnamefont{F.}~\bibnamefont{Pincet}},\ }%
  \bibfield{journal}{%
  \Doi{10.1103/PhysRevE.66.060501}{\bibinfo {journal} {Phys. Rev. E}}\ }%
  \textbf{\bibinfo {volume} {66}},\ \bibinfo {pages} {060501} (\bibinfo {year}
  {2002})%
  \bibAnnoteFile{NoStop}{bryanthardsphere}%
\bibitem{germain04}%
  \BibitemOpen
  \bibfield{author}{%
  \bibinfo {author} {\bibfnamefont{P.}~\bibnamefont{Germain}}, \bibinfo
  {author} {\bibfnamefont{J.~G.}\ \bibnamefont{Malherb}},\ and\ \bibinfo
  {author} {\bibfnamefont{S.}~\bibnamefont{Amokrane}},\ }%
  \bibfield{journal}{%
  \Doi{10.1103/PhysRevE.70.041409}{\bibinfo {journal} {Phys. Rev. E}}\ }%
  \textbf{\bibinfo {volume} {70}},\ \bibinfo {pages} {041409} (\bibinfo {year}
  {2004})%
  \bibAnnoteFile{NoStop}{germain04}%
\end{thebibliography}%
